\documentclass{aa}
\usepackage{txfonts}
\usepackage{graphicx}

\begin{document}

\title{The populations of planetary nebulae in the direction\\ 
       of the Galactic bulge
\thanks{Based on observations made at the South African Astronomical
        Observatory}
    }
\subtitle{Chemical abundances and Wolf-Rayet central stars}

\author{G\'{o}rny, S. K.
      \inst{1}
\and
      Stasi\'nska, G.
      \inst{2}
\and
      Escudero, A. V.
      \inst{3}
\and
      Costa, R. D. D.
      \inst{3}
}

\offprints{S. K. G\'{o}rny}

\institute{Copernicus Astronomical center, Rabia\'nska 8,
      87-100 Toru\'n, Poland \\
      \email{skg@ncac.torun.pl}
\and
      LUTH, Observatoire de Meudon, 5 Place Jules Janssen,
      F-92195 Meudon Cedex, France \\
      \email{grazyna.stasinska@obspm.fr}
\and
      Departamento de Astronomia, Instituto de Astronomia,
Geof\'{\i}sica e Ci\^encias Atmosf\'ericas da USP, Rua do Mat\~ao 1226,
05508-090,
                   S\~{a}o Paulo, Brazil
}

\date{Received ???; accepted ???}

\abstract{
We have observed 44 planetary nebulae (PNe) in the direction of the Galactic
bulge, and merged our data with published ones. We have distinguished, in
the merged sample of 164 PNe, those PNe most likely to prtain physically to
the Galactic bulge and those most likely to belong to the Galactic disk. We
have determined the chemical composition of all the 164 objects in a
coherent way. We looked for stellar emission features and discovered 14 new
[WR] stars and 15 new weak emission line central stars.

The analyzed data led us to the following conclusions: (1) The spectral type
distribution of [WR] stars is very different in the bulge and in the disk of
the Galaxy. However, the observed distributions are strongly dependent on
selection effects. (2) The proportion of [WR]\,PNe is significantly larger in
the bulge than in the disk. (3) The oxygen abundances in [WR] stars do no
appear to be significantly affected by nucleosynthesis and mixing in the
progenitors. (4) The O/H gradient of the Galactic disk PNe population
flattens in the most internal parts of the Galaxy. (5) The median oxygen
abundance in the bulge PN population is larger by 0.2~dex than in the disk
population seen in the direction of the bulge. (6) Bulge PNe with smaller
O/H tend to have smaller radial velocities. (7) The oxygen abundance
distribution of bulge PNe is similar in shape to that of the metallicity
distribution of bulge giants, but significantly narrower. (8) The location
of \object{SB 32} (\object{PN G 349.7-09.1}) in the ($V_{\rm{lsr}}$,
$l_{\rm{II}}$) diagram and its low oxygen abundance argues that it probably
belongs to the halo population.

}

\maketitle

\section{Introduction}

Planetary nebulae (PNe) are indicators of the chemical composition of the
interstellar matter out of which the progenitors stars were born (mainly
through the abundances of elements such as O, Ne, S). They are also
indicators of the nucleosynthesis and mixing processes occurring in their
central stars (mostly through the abundances of He, C, N).

Among planetary nebulae, those with central stars of Wolf-Rayet type have
received considerable attention recently (e.g. Tylenda et al. \cite{TAS93},
G\'{o}rny \& Stasi\'{n}ska \cite{GS95}, Pe\~{n}a et al. 2001, Bl\"ocker et
al. \cite{BOW01}, De~Marco \cite{dM03}, Acker \& Neiner \cite{AN03}).
Planetary nebulae with Wolf-Rayet central stars ([WR]\,PNe) are believed to
represent around 6-10\% of the total population of planetary nebulae
(G\'orny \& Stasi\'nska \cite{GS95}, Tylenda \cite{Ty96}). Until recently,
their mechanism of production was not well understood: how to obtain a PN
nucleus whose atmosphere is almost completely devoid of hydrogen and
essentially composed of helium and carbon? The late-helium flash scenario of
Iben et~al. (\cite{IKT83}) could possibly account for only a minority of
[WR]\,PNe, primarily those with an old nebular envelope surrounding a star of
late [WC] type. Acker et al. (\cite{AGC96}) and G\'orny \& Tylenda
(\cite{GT00}) have shown that the majority of [WR]\,PNe seem to draw an
evolutionary sequence from late-type to early-type [WC] types. Recently,
Herwig (\cite{He01}) and Bl\"ocker et al. (\cite{BOW01}) showed that
overshoot applied to models of stars in the AGB phase was able to produce
abundances similar to those observed in [WR] stars, and that a final thermal
pulse occurring on the AGB was able to produce [WR]\,PNe with young kinematic
ages. However, De~Marco \& Soker (\cite{dMS02}) make the point that the
double dust chemistry in [WR]\,PNe revealed by ISO spectra requires a
different explanation and propose a scenario where a low-mass star or planet
spirals into the AGB star, enhancing the mass-loss rate and introducing
extra mixing.

It is not clear yet whether [WR]\,PNe are produced only by a certain subclass
of intermediate mass stars. Neither the morphology, nor the chemical
composition, nor the Galactic distribution or kinematics of [WR]\,PNe seem to
differ statistically from those of non-[WR]\,PNe (G\'orny \& Stasi\'nska
\cite{GS95}, G\'orny \cite{Go01}). However, this statement needs to be
reexamined using a better defined control sample of non-[WR]\,PNe.

A coherent analysis of a large sample of planetary nebulae in the direction
of Galactic bulge offers the possibility of a more detailed comparison. One
of the advantages of planetary nebulae in the direction of the galactic
bulge is that their distances are known (to within about 10\% at least for
the majority of them). However, the population of planetary nebulae in the
Galactic bulge is probably not the same as that of the Galactic disk, since
the stellar population in the bulge is 10 Gyr old (Zoccali et al. 
\cite{ZRO03}), while in the disk the stellar ages span the entire range from
10 Gyr to virtually zero (Chiappini et al. \cite{ChMR01}). Therefore, it is
also interesting to investigate whether the global properties of [WR]\,PNe in
the Galactic bulge are different from those of disk [WR]\,PNe. In this
respect, it is remarkable that in the Magellanic Clouds, the spectral types
of [WR]\,PNe central stars lie between [WC4] and [WC8] (Pe\~na et al.
\cite{PHK97}) (except the nucleus of \object{N66}, which until recently was
the only known [WN] type star until another case was found in our Galaxy by
Parker \& Morgan \cite{PM03}) while the [WR] PNe known in the Galaxy are
either of very late ([WC8] - [WC11]) or very early ([WC2] - [WC4]) types
(G\'orny \cite{Go01}, Acker \& Neiner \cite{AN03}), clearly avoiding the
middle types for some unknown reason.  This suggests that the WR phenomenon
is dependent on the charateristics of the stellar population giving rise to
the observed PNe. For the Galactic bulge, it has also been proposed that
[WR] stars are mostly of intermediate spectral types (G\'orny \cite{Go01}),
although Acker \& Neiner (\cite{AN03}) do not confirm such a view. However
all these inferences are based on small samples.

In this paper, we first enlarge the sample of [WR]\,PNe towards the Galactic
bulge, by presenting a new set of observations. We combine our observations
with those of previous data sets which also allow us to detect the possible
presence of stellar emission lines and to determine which of these PNe
harbour a [WR] type central star (or weak emission line star (WELS) possibly
related to [WR] stars, see e.g. Tylenda et al. \cite{TAS93}, Pe\~na et al.
\cite{PSM01}). The samples are described in Sect. 2, and rearranged into one
sample of PNe pertaining physically to the Galactic bulge and another sample
of PNe seen in the direction of the Galactic bulge but pertaining to the
Galactic disk. In Sect. 3, we determine the chemical composition of all the
PNe in the combined sample and discuss the accuracy of the derived
abundances. In Sect. 4, we show the results of our search of [WR] stars and
WELS in the entire sample. We define the spectral types of the newly
discovered [WR] stars, and comment on the statistics of the [WC] types as
well as on the physical location of the [WR]\,PNe, including a discussion of
the observational selection effects. In Sect. 5, we comment on the chemical
composition of the PNe, by comparing several subsamples. Our main
conclusions are summarized in Sect. 6.

\section{Observational data}

\subsection{New observations and reduction procedure}

The main aim of our observations was to detect new Wolf-Rayet central stars
in the direction of the Galactic center (i.e. within, say, 20 degrees from
the Galactic center in Galactic longitude and 15 degrees in Galactic
latitude). The selection of PNe for observations was based on two criteria.
First, we have dismissed all PNe for which there already existed spectral
information on the central star, i.e. the spectral type was known or the
stellar continuum had been observed but no emission lines were seen (Tylenda
et al. \cite{TAS93}, Tylenda et al. \cite{TAS91}). Next, we selected objects
with IRAS mid-infrared colours similar to the colours of known [WR]\,PNe (see
G\'orny et al. \cite{GSS01}). Therefore, obviously, this sample cannot be
used to estimate the proportion of [WR]\,PNe among PNe in the Galactic bulge.
For comparison purposes and to assess the quality of our spectra we also
observed a few objects already known to have emission line central stars.

The spectra of the 44 objects of our observational sample were obtained in
July 2000 with the 1.9-meter telescope at the South African Astronomical
Observatory. The spectral coverage was 3500-7000\AA\ with an average
resolution of 1000. A slit width of 1.8 arcsec was selected, resulting in
some loss of stellar light due to seeing conditions but allowing to avoid
blending of important nebular lines. The typical integration time was 60
minutes (30 minutes if the central star was a known [WR] or WELS) with
additional short time exposure to secure unsaturated spectra of the
strongest nebular lines. The log of the observations is presented in Table
1\footnote{Table 1 is available in electronic form only.}. Two spectroscopic
standard stars, \object{Feige\,110} and \object{CD\,-32\,9927}, were also
observed each night. A CuAr lamp exposure was taken before or after each PN
spectrum for wavelength calibration purposes.

The reduction of the spectra was done with standard procedures of the
MIDAS{\footnote{MIDAS is developed and maintained by the European Southern
Observatory.}} (99NOV) longslit spectra package. These included bias
subtraction, flat-field correction, atmospheric extinction correction,
wavelength and flux calibration and extraction of the 1-dimensional spectra.
The line fluxes were derived with the REWIA package (J. Borkowski;
http://www.ncac.torun.pl/$^\sim$jubork) adopting Gaussian profiles and
performing multi-Gaussian fits in case of blended lines.

% figure 1
\begin{figure*}
\resizebox{0.88\hsize}{!}{\includegraphics{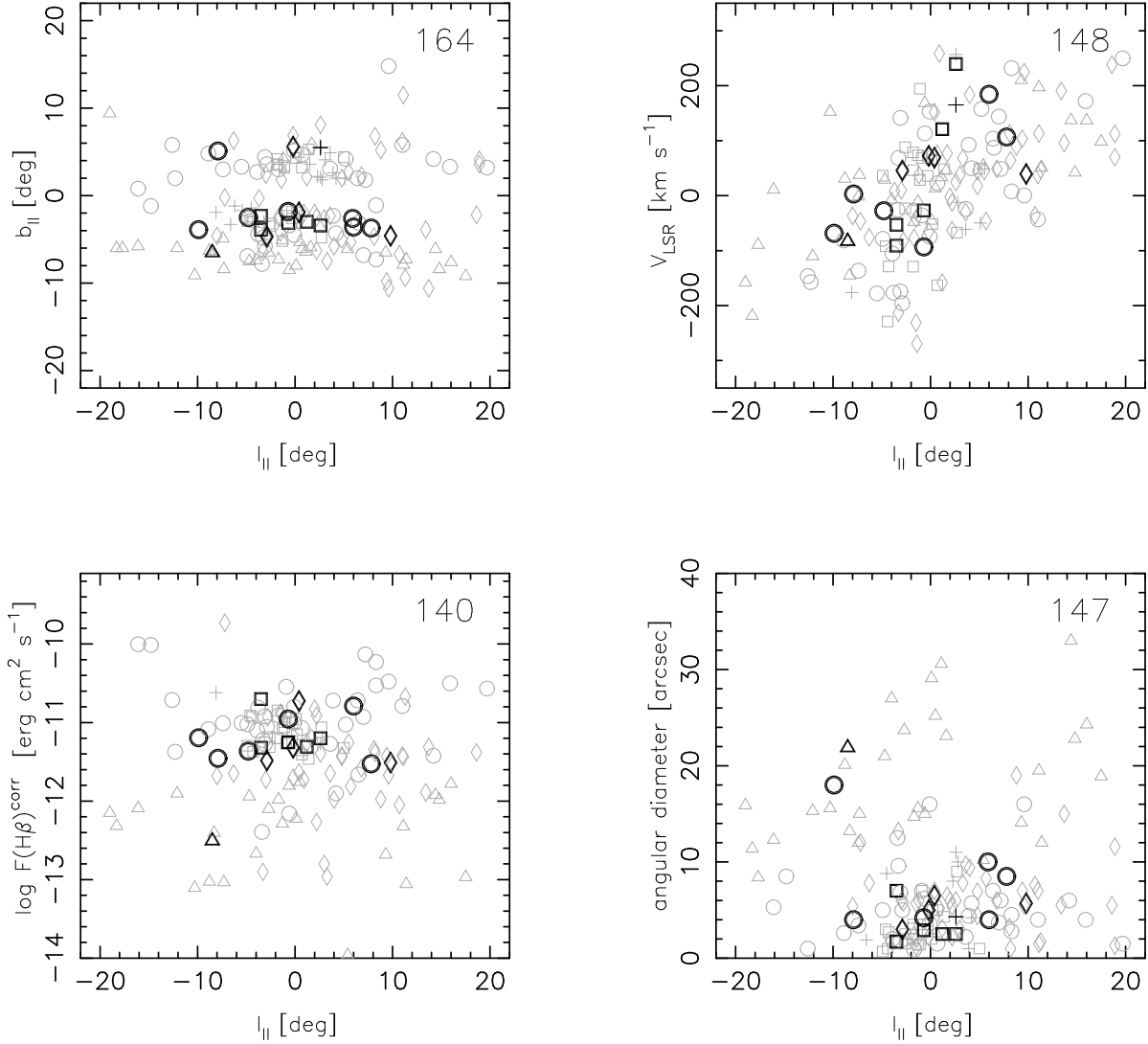}}
\caption[]{
Distribution in various diagrams of the PNe from our merged sample. Top
left: $b_{\rm{II}}$ vs. $l_{\rm{II}}$; top right: $V_{\rm{lsr}}$, the radial
velocities corrected for solar motion vs. $l_{\rm{II}}$; bottom left: total
nebular flux in H$\beta$ corrected for reddening vs. $l_{\rm{II}}$; bottom
right: angular diameters as a function of $l_{\rm{II}}$. The number on the
top right of each panel gives the total number of objects represented in the
plot. Objects pertaining to different samples are marked by different
symbols: G: circles, C: squares, B: triangles, K: plus signs, E: diamonds.
(See Sect. 2.2 for detailed definitions.) Symbols for [WR]\,PNe are thicker
than for the rest of PNe.
}
\label{f1}
\end{figure*}

\subsection{A merged sample of PNe in the direction of the Galactic bulge}

Published spectroscopic data on planetary nebulae towards the Galactic bulge
are numerous (see Stasi\'nska et al. \cite{SRMcC98} for a compilation until
1997). However, the only data that allow us to check for the presence of
emission lines in the central stars are those of Cuisinier et al.
(\cite{CMK00}) (30 objects), Escudero \& Costa (\cite{EC01}) (45 objects)
and Escudero et al. (\cite{ECM04}) (57 objects). We thus merged these three
samples with the one described in Sect. 2.1, obtaining a total of 164
objects composing what we will refer to as the merged sample. In the case of
duplicate spectra, we kept for further analysis the ones described in Sect.
2.1.

Fig.~\ref{f1} shows  the distribution in various diagrams of the PNe from our
merged sample. Each of the set of observations is marked by a different
symbol: circles are for subset G (the set of observations described in Sect.
2.1), squares are for subset C (the sample of Cuisinier et al. \cite{CMK00}),
triangles are for subset B (the objects from the list of Beaulieu et al.
\cite{BDF99} observed by Escudero \& Costa \cite{EC01}), plus signs for
subset K (the objects from the list of Kohoutek \cite{Ko94} observed by
Escudero \& Costa \cite{EC01}) and diamonds for subset E (the objects
observed by Escudero et al. \cite{ECM04}). [WR]\,PNe are marked by thicker
symbols.  The top left panel shows the Galactic latitude and longitude,
$b_{\rm{II}}$ and $l_{\rm{II}}$. The top right panel shows $V_{\rm{lsr}}$,
the radial velocities corrected for solar motion as a function of
$l_{\rm{II}}$ (the radial velocities have been taken from Durand et al. 
\cite{DAZ98} and have been corrected from solar motion using the formulae
given in Beaulieu et al. \cite{BFK00}). The bottom left panel shows the
total extinction corrected nebular flux in H$\beta$ 
   \footnote{The H$\beta$ fluxes for most of objects come from Acker et al.
(\cite{AOS92}) and have been corrected for interstellar extinction as
determined from the ratio of the radio to H$\beta$ fluxes if good quality
radio flux measurements are available (references as in Stasi\'nska et al.
\cite{STA92}). Otherwise the extinction has been derived from the
H$\beta$/H$\alpha$ ratio using values from Tylenda et al. (\cite{TSA94}).
For objects from the list of Beaulieu et al. (\cite{BDF99}) we have
calculated the H$\beta$ fluxes from H$\alpha$ fluxes given in this paper
using the H$\beta$/H$\alpha$ ratios from our spectra to calculate the
extinction.}
    as a function of $l_{\rm{II}}$. The bottom right panel shows the angular
diameters as a function of $l_{\rm{II}}$. A clear segregation of the various
subsamples is seen. This is due to the various selection criteria used to
define each sample. The sample of Cuisinier et al. (\cite{CMK00}) is part of
the list of PNe from Stasi\'nska \& Tylenda (\cite{ST94}) that lie within 10
degrees of the Galactic center, have radio fluxes and diameters measured
with the VLA, with angular diameters smaller than 15 arcsec and radio fluxes
at 6 cm between 10 and 100 mJy. The PNe from samples G and B extend over
larger zones in $l_{\rm{II}}$ and $b_{\rm{II}}$. But, while the objects from
sample G are rather luminous, the objects from sample B are newly discovered
PNe that are rather faint and have larger angular diameters. The objects
from sample E are found in a more extended zone than the objects from sample
C and are of intermediate luminosity. Since these different observational
sets obey to different selection criteria, none of them can be considered to
represent the population of the Galactic bulge and it is no surprise that
each one may have different mean properties (as already noted by Escudero \&
Costa \cite{EC01}). By merging these data sets, we are getting a more
complete view of the PN populations in the direction of the Galactic bulge,
although, unfortunately, none of the samples we consider is complete in any
sense.

We can divide our merged sample of PNe into two distinct populations. The
first one, hereafter referred to as $b$, is composed of objects lying within
10 degrees of the Galactic center, having angular diameters smaller than 20
arcsec and radio fluxes at 6 cm smaller than 100 mJy. About 95\% of these
objects are likely physically related to the Galactic bulge as shown by
Stasi\'nska et al. (\cite{STA91}). The second subsample, hereafter referred
to as $d$, contains the remaining objects and most of them should be related
to the Galactic disk population.

Note that this distinction between "bulge" and "disk" populations does not
use the radial velocities of the nebulae.  However, as already noted by
Zijlstra et al. (\cite{ZAW97}), PNe thought to be physically related to the
Galactic bulge exhibit a large range of radial velocities, between
-250~km~s$^{-1}$ and +250~km~s$^{-1}$. The distinction in kinematic
properties between our subsamples $b$ and $d$ of PNe in the direction of the
Galactic bulge is clearly seen by comparing Figs.~\ref{f12} and~\ref{f13}
(to be fully described later), which repeat the ($V_{\rm{lsr}}$,
$l_{\rm{II}}$) diagram for samples $b$ and $d$ respectively. Objects from
sample $d$ that are found at a Galactic longitude close to zero have a
radial velocity close to zero, compatible with rotation in circular orbit in
the Galactic disk.

\section{Analysis of nebular emission lines}

The chemical composition of the newly observed planetary nebulae was derived
as described below.  We also rederived the abundances of the PNe with
published line intensities using the same methods in order to obtain a
reasonably homogeneous set of abundances.

\subsection{Reddening correction}

The measured line intensities of the newly observed sample have first been
corrected for interstellar extinction using the extinction law of Seaton
(\cite{Se79}) and forcing the observed H$\alpha$/H$\beta$ ratio to the
theoretical recombination value. However, with such a procedure, the
reddening-corrected H$\delta$/H$\beta$ and H$\gamma$/H$\beta$ ratios were
often significantly different than theoretically expected. We attribute this
to the E-W orientation of the slit and to the fact that the spectra were
taken at different airmasses (reaching values of 2) which resulted in
substantial loss of registered light at short wavelengths. It is important,
for abundance determinations, to use intensities of [O~{\sc iii}]
$\lambda$4363 and [O~{\sc ii}] $\lambda$3727 as reliable as possible.
Therefore we have used the H$\delta$/H$\beta$ and H$\gamma$/H$\beta$ ratios
and have calculated the values of the additional correction that would bring
them to the teoretical recombination values. This correction was then
linearly propagated to the observed fluxes at all wavelengths shorter of
4681\AA (H$\beta$).  The resulting line fluxes are presented in Table
2\footnote{Table 2 is available in electronic form only.} on the scale of
H$\beta$=100. We also give line identification and laboratory wavelengths.

The same additional correction procedure has been applied to the objects
from Escudero \& Costa (\cite{EC01}) and Escudero et al. (\cite{ECM04})
samples that originally also revealed some deviations from the theoretical
Balmer decrement as discussed in Escudero et al. (\cite{ECM04}).

\subsection{Nebular abundance determinations}

% figure 2
\begin{figure}
\resizebox{0.9\hsize}{!}{\includegraphics{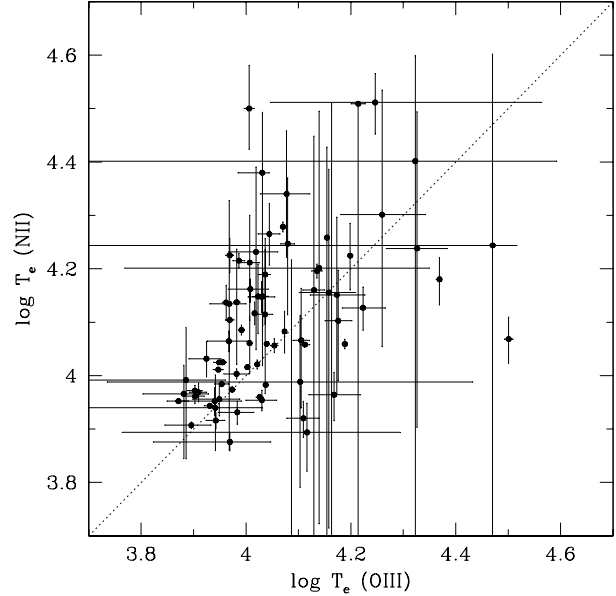}}
\caption[]{
The electron temperature derived from [O~{\sc iii}] $\lambda$4363/5007
versus the electron temperature derived from [N~{\sc ii}] $\lambda$5755/6584.
}
\label{f2}
\end{figure}

% figure 3
\begin{figure}
\resizebox{0.9\hsize}{!}{\includegraphics{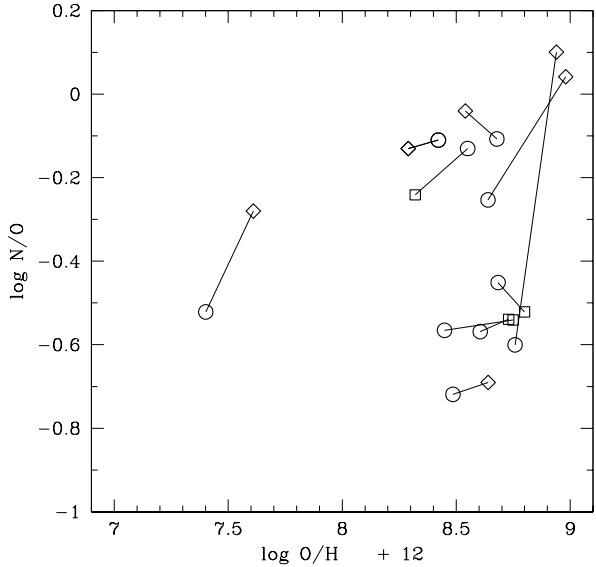}}
\caption[]{
The values of log N/O versus log O/H +12 for all the objects in common in
the various data sets composing or merged sample. Symbols corresponding to
the same object are linked by thin lines. The meaning of the symbols is the
same as in Fig.~\ref{f1}.
}
\label{f3}
\end{figure}

The abundances were derived with the classical empirical method.  The
electron densities were obtained from the [S~{\sc ii}] $\lambda$6731/6717
ratio, the electron temperatures from the [O~{\sc iii}] $\lambda$4363/5007
and [N~{\sc ii}] $\lambda$5755/6584 ratios. The ionic abundances and
elemental abundances were then derived using the code ABELION as in Pe\~na
et al. (\cite{PSM01}). The only difference is that here we used the
collision strengths of Keenan et al. (\cite{KAB96}) for [S~{\sc ii}] lines.
Those objects for which the electron temperature could be determined neither
from [O~{\sc iii}] $\lambda$4363/5007 nor from [N~{\sc ii}]
$\lambda$5755/6584 were discarded from abundance determinations. This
concerns only 5 objects among 164 (4 in sample $b$, one in sample $d$). The
temperature derived from [N~{\sc ii}] was used for ions of low ionisation
potential, that from [O~{\sc iii}] $\lambda$4363/5007 for ions of high
ionization potential. In general, the temperatures from [O~{\sc iii}]
$\lambda$4363/5007 and [N~{\sc ii}] $\lambda$5755/6584 are similar, but
often not identical within their error bars as shown in Fig.~\ref{f2}. If
the temperature from [N~{\sc ii}] was very uncertain, we rather used the
temperature from [O~{\sc iii}] $\lambda$4363/5007 for all the ions. We have
adopted for the abundance of O$^{+}$ the mean of the abundances derived from
the [O~{\sc ii}] $\lambda$3727 and the [O\,{\sc{ii}}]\,$\lambda$7320,7330
line, when available. A discussion of the comparison between both abundances
can be found in Escudero et al. (\cite{ECM04}). Table 3
   \footnote{Table 3 is available in electronic form only.} 
lists the resulting plasma diagnostics and abundances for all the objects of
the merged sample. Column (1) gives the PNG number; Column (2) gives the
usual name of the PN; Column (3) indicates the observational subsample (G,
C, B, K or E). Column (4) indicates whether the object belongs to sample "b"
as defined in Sect. 2.2. Column (6) gives the electron density deduced from
[S~{\sc ii}] $\lambda$6731/6717, columns (7) and (8) give the electron
temperature deduced from [N~{\sc ii}] $\lambda$5755/6584 and [O~{\sc iii}]
$\lambda$4363/5007 respectively. Columns (9) gives the He/H ratio, column
(10), (11), (12), (13) give the N/H, O/H, Ne/H, S/H ratios, respectively, in
units of 10$^{-6}$. Columns (14),(15),(16),(17), (18) give He$^{+}$/H$^{+}$,
He$^{++}$/H$^{+}$, O$^{+}$/H$^{+}$ as derived from the 3727 line,
O$^{+}$/H$^{+}$ as derived from the 7325 line, O$^{++}$/H$^{+}$. Column (19)
gives the logarithmic extinction at H$\beta$ derived from the spectra. In
this table, there are three rows for each object, and a fourth row used to
separate the objects. The first row gives the values of parameters computed
from the nominal values of the observational data. The second and third row
give the upper and lower limit respectively of these parameters as derived
taking into account the observational error bars.

Errors in the derived abundances result not only from observational errors
(including the dereddening procedures), but also from uncertainties in the
atomic data and the adequacy of the abundance derivation scheme. Comparing
abundances derived by different authors for the same objects gives an idea
of the uncertainty including all the sources of errors. This is shown in
Fig.~\ref{f3}, where we plot the values of log N/O versus log O/H +12 for
all the objects in common among several of the subsamples we considered.
Values pertaining to the same object are linked by thin lines.  The symbols
are the same as in Fig.~\ref{f1}. From this figure, we conclude that global
uncertainties in abundance ratios are typically of 0.2 -- 0.3 dex for O/H
but can be much larger in some cases for N/O. Apart from different
determinations of the temperatures, the main source of uncertainty in the
oxygen abundance is the estimation of O$^{+}$/H$^{+}$, especially in low
excitation objects. The [O~{\sc ii}] $\lambda$3727 line is severely affected
by reddening, while the [O\,{\sc{ii}}]\,$\lambda$7320,7330 line is
intrinsically much weaker, more dependent on temperature and strongly
affected by possible recombination. Both lines are sensitive to density, but
in different ways. The N/O ratio is less affected by errors in the electron
temperature than the O/H ratio. On the other hand, its derived value
strongly depends on the value adopted for the O$^{+}$ abundance. Quite
often, a difference in N/O between the determination by two authors traces
back to a different estimate of the electron density, which has important
consequences for very dense nebulae.

For helium, the expected errors in the abundances are much smaller, except
in the case of PNe of very low excitation, where no correction can be made
for the presence of neutral helium. However, even for high excitation
nebulae, the errors are relatively large compared to expected
object-to-object variations.

% figure 4
\begin{figure}
\resizebox{0.9\hsize}{!}{\includegraphics{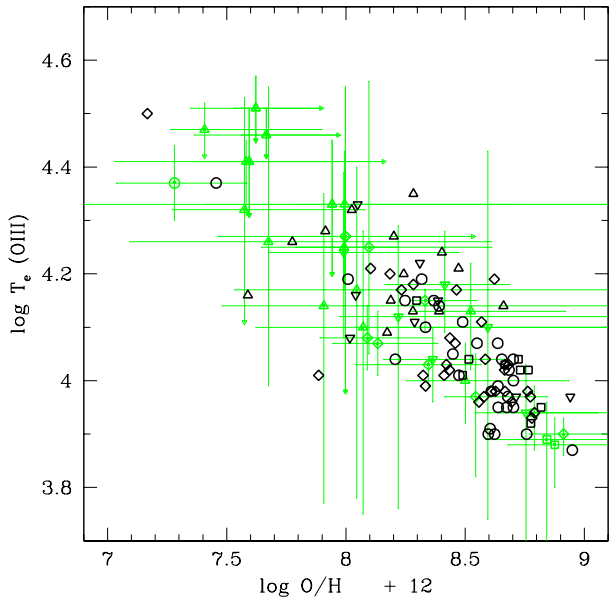}}
\caption[]{
The value of the electron temperature $T_e$ as derived from [O~{\sc iii}]
$\lambda$4363/5007 as a function of the computed value of log O/H + 12. The
observational error bar on $T_e$ and the error bar on log O/H resulting from
the error in $T_e$ is shown (unless the error in log $T_e$ is smaller than
0.1 and the error in log O/H smaller than 0.2, in which case the symbol is
marked with a black heavy line). Arrows indicate that only an upper limit on
$T_e$ or a lower limit on O/H ca be obtained. The shape of the symbols has
the same meaning as in Fig.~\ref{f1}.
}
\label{f4}
\end{figure}

The lowest oxygen abundances found in our merged sample are as low as log
O/H +12 = 7.1 -- 7.2. One may suspect that such low values are not real but
result simply from an overestimated electron temperature (which, in general,
is the main source of uncertainty in the oxygen abundance derivation).  In
Fig.~\ref{f4}, we plot the value of the electron temperature $T_e$, as
derived from [O~{\sc iii}] $\lambda$4363/5007, as a function of the computed
value of log O/H + 12, together with the observational error bar on $T_e$
and the error bars on log O/H resulting only from $T_e$. This figure shows
that indeed, some of the lowest derived abundances are affected by large
error bars, implying that the true abundances might be significantly higher
than derived. However, there are four objects, for which log O/H + 12 is
definitely smaller than 7.6. We are, of course, aware of the fact that
recombination lines in planetary nebulae lead to larger values of the
elemental abundances, sometimes by factors up to 10 (see e.g. Liu
\cite{Li02} for a review). Here we adopt the view, as in all statistical
studies on abundances so far, that the abundances derived from the forbidden
lines are representative of the chemical composition in the bulk of the
nebula.

\section{[WR]\,PNe}

\subsection {Search for new [WR]\,PNe}

We first searched our new spectra for the presence of emission line central
stars.  The distinction between [WR] stars and WELS is made as in Tylenda et
al. (\cite{TAS93}) on the basis of the strength and width of the C~{\sc iv}
$\lambda$5805 doublet.  This doublet is often seen as two separate
components in WELS and is much weaker and narrower than in [WR]-type stars.
It is the only stellar emission line seen in the spectra of WELS, among the
lines used to classify [WR] stars.  Note that the difference between [WR]
stars and WELS is not always easy to make and one may suspect that better
quality spectra could perhaps reveal additional spectral features especially
for cooler stars.

We found 4 new [WR] stars and 8 new WELS in sample G, in addition to the 3
[WR] stars and 8 WELS already known to exist in this sample.

We also examined the spectra of  Escudero \& Costa (\cite{EC01}) and
Escudero et al. (\cite{ECM04}). In the Escudero \& Costa\cite{EC01})
subsample we found 2 [WR] stars and 3 WELS among 45 PNe. In the Escudero et
al. (\cite{ECM04}) sample we found 3 new [WR] stars and 4 new WELS in
addition to the 1 known [WR] and 1 known WELS.

The observations by Cuisinier et al. (\cite{CMK00}) are published only as a
list of emission lines, among which lines of C~{\sc iii} $\lambda$5695 and
C~{\sc iv} $\lambda$5805 appear. These lines are in fact of stellar origin
and we attribute them to [WR] stars. With these criteria, we find 5 new
[WR]\,PNe in the sample of Cuisinier et al. (\cite{CMK00}).

In total, including the [WR] and WELS already known, we find 18 [WR] and 24
WELS among 164 objects considered. The rate of occurence of [WR]\,PNe in the
direction of the bulge is thus somewhat higher than that found by Tylenda et
al. (\cite{TAS93}) by examining 350 PNe spectra in the whole Galaxy. The
ratio of WELS is much higher. This is likely due to two reasons. First, the
spectra we examined are of substantially better quality than those examined
by Tylenda et al. (\cite{TAS93}), which came from a snapshot spectroscopic
survey of the Galaxy. Second, our own observations were conducted on a
sample specially designed to favor new discoveries of [WR]\,PNe, as explained
in Sect. 2.1.

In the following of this paper, we will concentrate on [WR]\,PNe and their
comparison with remaining PNe.

Table 4 presents all the PNe in the direction of the Galactic bulge whose
nuclei were found to present emission lines. The [WR] stars and WELS
discovered by us constitute "group 1". The [WR] stars and WELS that were
already known to exist in the samples we analyzed constitute "group 2".
There are also 15 [WR]\,PNe and 11 WELS reported in the literature to exist
in the direction of the Galactic bulge but that do not belong to our
"merged" sample\footnote { Recently, 2 additional [WR]\,PNe were discovered
in the direction of the Galactic bulge (Parker \& Morgan \cite{PM03}) during
a follow-up spectroscopic survey of 1000 new PNe candidates proposed from
visual scans of AAO/UKST narrow-band H$\alpha$ survey of the Milky Way
(Parker et al. \cite{PPM99}, Parker et al. \cite{PHR03}).  However, among
700 objects observed to date, only 7 new [WR]\,PNe were found (Parker \&
Morgan \cite{PM03}). This very low detection rate possibly involves
important selection effects and these new objects have to be considered
separately in future works.}. They are also listed in Table 4 for
completeness, under "group 3". For each of these groups, column (1) gives
the PN\,G number; column (2) gives the usual name of the PN; column (3)
indicates the observational subset (using the same convention as in column
(3) of Table 3); column (4) indicates whether the object belongs to sample
$b$ as defined in Sect. 2.2; column (5) gives the FWHM of the C~{\sc iv}
$\lambda$5805 or C~{\sc iii} $\lambda$5695 line; column (6) gives the
category of the star: W for [WR], w for WELS. For convenience, the same
indications (i.e. W or w), are reported also in column (5) of Table 3.

\addtocounter{table}{3}

\begin{table*}
\caption{[WR] and WELS planetary nebulae in the direction of the Galactic
            bulge.}
\begin{flushleft}
{\scriptsize
\begin{tabular}{
                    l
                    l @{\hspace{0.10cm}}
                    c @{\hspace{0.10cm}}
                    c @{\hspace{0.10cm}}
                    c @{\hspace{0.20cm}}
                    c @{\hspace{0.30cm}}
                    c
                    l @{\hspace{0.10cm}}
                    c @{\hspace{0.10cm}} }
\hline
    PN G &
    name &
    sample &
    Bulge &
    FWHM &
    W/w &
    type &
    spectra details or reference &
    Acker\&Neiner (2003) \\
  ~~~1 & ~~~2 & 3 & 4 & 5 & 6 & 7 & ~~~8 & 9 \\
\hline
              & & & & & & & & \\
\multicolumn{9}{l}{ group 1:  [WR]\,PNe discovered in this work } \\
    000.4+04.4 & \object{K 5- 1}       & E     & b & 16.0        & w & wels     & & \\
    000.7+04.7 & \object{H 2-11}       & G$^2$ & b & 18.0        & w & wels     & & \\
    000.9-02.0 & \object{Bl 3-13}      & E     & b & 7.8+7.4     & w & wels     & & \\
    001.2-03.0 & \object{H 1-47}       & C     & b &             & W & [WC11]?  & CIII present, CIV absent & (? - very late) \\
    002.6-03.4 & \object{M 1-37}       & C     & b &             & W & [WC11]?  & CIII present, CIV absent & (? - very late) \\
    002.6+05.5 & \object{K 5- 3} $^1$  & K     & b & 28.2        & W & [WC4]    & CIV strong, OV$>$OVI, OVI 3822 present &  ([WO3]) \\
    003.6+03.1 & \object{M 2-14}       & G     & b & 21.8        & w & wels     & & \\
    005.9-02.6 & \object{MaC 1-10}     & G     & b & 37.4        & W & [WC8]    & CIV $\approx$ CIII & ([WC7-8])\\
    006.5-03.1 & \object{H 1-61}       & G     & b & 12.6+11.2   & w & wels     & & \\
    008.3-01.1 & \object{M 1-40}       & G     &   & instr+instr & w & wels     & & \\
    008.1-04.7 & \object{M 2-39}       & E     & b & 19.4        & w & wels     & & \\
    008.3-07.3 & \object{NGC 6644}     & G     &   & 6.0+6.8     & w & wels     & & \\
    009.6+14.8 & \object{NGC 6309}     & G     &   & 8.7+7.3     & w & wels     & & \\
    009.8-04.6 & \object{H 1-67}       & E     &   & 30.3        & W & [WC2-3]  & CIV present, OVI5290 ?, OVI3822 strong & ([WO2]?)\\
    009.6-10.6 & \object{M 3-33}       & E     &   & 7.6+7.8     & w & wels     & & \\
    013.7-10.6 & \object{Y-C 2-32}     & E     &   & 7.3+6.7     & w & wels     & & \\
    014.4-06.1 & \object{SB 19}        & B     &   & 15.3        & w & wels     & & \\
    343.9-05.8 & \object{SB 30}        & B     &   & 9.2         & w & wels     & & \\
    347.4+05.8 & \object{H 1- 2}       & G     &   & instr+instr & w & wels     & & \\
    350.1-03.9 & \object{H 1-26}       & G     &   & 34.8        & W & [WC4-5]  & CIV$>>$CIII, OV$>$OVI \& OV$>$CIII &  ([WO4]) \\
    351.5-06.5 & \object{SB 34}        & B     &   & 31.6        & W & [WC2]    & CIV present, OVI\,3822 very strong, & ([WO2]) \\
               &                       &       &   &             &   &          & OVI\,5290 ?, OVII\,5670 ? \\
    351.7-06.6 & \object{SB 35}        & B     &   & instr+instr & w & wels     & & \\
    352.1+05.1 & \object{M 2- 8}       & G$^2$ & b & 31.8        & W & [WC2-3]  & OVI5290 present, OVI3822 present & ([WO3]) \\
    356.1+02.7 & \object{Th 3-13}      & G     & b & 21.5:       & w & wels     & & \\
    356.5-02.3 & \object{M 1-27}       & C     & b &             & W & [WC11]?  & CIII present, CIV absent & (? - very late) \\
    356.5-03.9 & \object{H 1-39}       & C     & b &             & W & [WC11]?  & CIII present, CIV absent & (? - very late)\\
    357.1-04.7 & \object{H 1-43}       & E     & b & instr $^3$  & W & [WC11]   & CIII present, CIV absent, CII ? & (? - very late) \\
    359.8+05.6 & \object{M 2-12}       & E     & b & instr $^3$  & W & [WC11]   & CIII present, CIV absent, CII present & (? - very late) \\
    359.3-01.8 & \object{M 3-44}       & G     & b & instr $^3$  & W & [WC11]   & CIII present, CIV absent, CII ? & (? - very late) \\
    359.3-03.1 & \object{M 3-17}       & C     & b &             & W & [WC11]?  & CIII present, CIV absent & (? - very late) \\
               & & & & & & & & \\
\multicolumn{9}{l}{ group 2:  known [WR]\,PNe belonging to our "merged sample"} \\
    000.4-01.9 & \object{M 2-20}       & E     & b & 28.0        & W & [WC5-6]? & 10* & [WC5-6] \\
    002.6+08.1 & \object{H 1-11}       & E     & b & instr+instr & w & wels     & & \\
    004.2-04.3 & \object{H 1-60}       & G     & b & instr+instr & w & wels     & & \\
    006.4+02.0 & \object{M 1-31}       & G     & b & 25.3        & w & wels     & & \\
    006.0-03.6 & \object{M 2-31}       & G     & b & 41.8        & W & [WC4-5]? & 10* & [WC4] \\
    007.8-03.7 & \object{M 2-34}       & G     & b & no emission & W & [WC]     & 1   & \\
    007.0-06.8 & \object{VY 2- 1}      & G     & b & instr+instr & w & wels     & & \\
    351.1+04.8 & \object{M 1-19}       & G     &   & 17.7        & w & wels     & & \\
    355.2-02.5 & \object{H 1-29}       & G     & b & 34.3        & W & [WC4-5]? & 10* & [WC4] \\
    356.2-04.4 & \object{Cn 2-1}       & G     & b & no emission & w & wels     & & \\
    356.7-04.8 & \object{H 1-41}       & G     & b & instr+instr & w & wels     & & \\
    357.1+03.6 & \object{M 3- 7}       & G     & b & 16.2        & w & wels     & & \\
    359.9-04.5 & \object{M 2-27}       & G     & b & 22.5        & w & wels     & & \\
               & & & & & & & & \\
\multicolumn{9}{l}{ group 3:    known [WR]\,PNe not belonging to our "merged sample"} \\
    001.5-06.7 & \object{SwSt 1}       &       &   & & W & [WC10]   & 6*,8*,10*  & [WC9]pec \\
    002.4+05.8 & \object{NGC 6369}     &       &   & & W & [WC4]    & 1,3,10*    & [WO3] \\
    002.0-06.2 & \object{M 2-33}       &       & b & & w & wels     & 9          &  \\
    002.2-09.4 & \object{Cn 1-5}       &       & b & & W & [WC4]    & 1,10*      & [WO4]pec \\
    002.0-13.4 & \object{IC 4776}      &       &   & & w & wels     & 9          &  \\
    003.1+02.9 & \object{Hb 4}         &       &   & & W & [WC4]    & 8*,10*     & [WO3] \\
    003.9-14.9 & \object{Hb 7}         &       &   & & w & wels     & 9          &  \\
    004.6+06.0 & \object{H 1-24}       &       & b & & w & wels     & 9          &  \\
    004.9+04.9 & \object{M 1-25}       &       & b & & W & [WC6]    & 2,10*      & [WC5-6] \\
    006.8+04.1 & \object{M 3-15}       &       & b & & W & [WC5]    & 10*        & [WC4] \\
    007.8-04.4 & \object{H 1-65}       &       & b & & w & wels     & 9          &  \\
    009.4-05.0 & \object{NGC 6629} $^4$&       &   & & W & [WC5-6]  & 7          & [WC4]? \\
    010.8-01.8 & \object{NGC 6578} $^4$&       &   & & W & [WC4-6]  & 7          &  \\
    011.9+04.2 & \object{M 1-32}       &       &   & & W & [WC4]    & 10*        & [WO4]pec \\
    011.7-00.6 & \object{NGC 6567} $^4$&       &   & & W & [WC5-6]  & 7          &  \\
    012.2+04.9 & \object{PM 1-188}     &       &   & & W & [WC10]   & 6*,5,10*   & [WC10] \\
    012.5-09.8 & \object{M 1-62}       &       &   & & w & wels     & 9          &  \\
    014.3-05.5 & \object{V-V 3-6}      &       &   & & w & wels     & 9          &  \\
    016.4-01.9 & \object{M 1-46}       &       &   & & w & wels     & 9          &  \\
    017.9-04.8 & \object{M 3-30}       &       &   & & W & [WC2-3]  & 10*        & [WO2] \\
    019.7-04.5 & \object{M 1-60}       &       &   & & W & [WC4]    & 2          & [WC4] \\
    019.4-05.3 & \object{M 1-61}       &       &   & & w & wels     & 9          &  \\
    341.8+05.4 & \object{NGC 6153}     &       &   & & w & wels     & 9          &  \\
    352.9+11.4 & \object{K 2-16}       &       &   & & W & [WC11]   & 1,4,8*,10* & [WC11] \\
    355.9-04.2 & \object{M 1-30}       &       & b & & w & wels     & 9          &  \\
    358.3-21.6 & \object{IC 1297}      &       &   & & W & [WC3-4]  & 1,10*      & [WO3] \\
\hline
\end{tabular}
}
\parbox{17.5cm}{
notes: 1) \object{K 5-3} = \object{IRAS17276-2342} in Acker et al. (\cite{AGC96});
       2) spectra of Escudero \& Costa (\cite{EC01}) (sample E) were used to
             measure stellar emissions;
       3) FWHM of C~{\sc iii} $\lambda$5695;
       4) objects classified as WELS in Pe\~na et al. (\cite{PSM01})
}

~

\parbox{17.5cm}{
references:  1)   Tylenda et al. (\cite{TAS93});
             2)   Acker et al. (\cite{AGC96});
             3)   Koesterke \& Hamann (\cite{KH97});
             4)   Leuenhagen et al. (\cite{LHJ96});
             5)   Leuenhagen \& Hamann (\cite{LH98});
             6*)  reclassified using data of Crowther et al. (\cite{CdMB98});
             7)   Pe\~na et al. (\cite{PSM01});
             8*)  reclassified using data of de Ara\'ujo et al. (\cite{dAMP02});
             9)   Acker \& Neiner (\cite{AN03});
             10*) reclassified using data of Acker \& Neiner (\cite{AN03});
}
\end{flushleft}
\end{table*}

% figure 5
\begin{figure}
\resizebox{0.81\hsize}{!}{\includegraphics{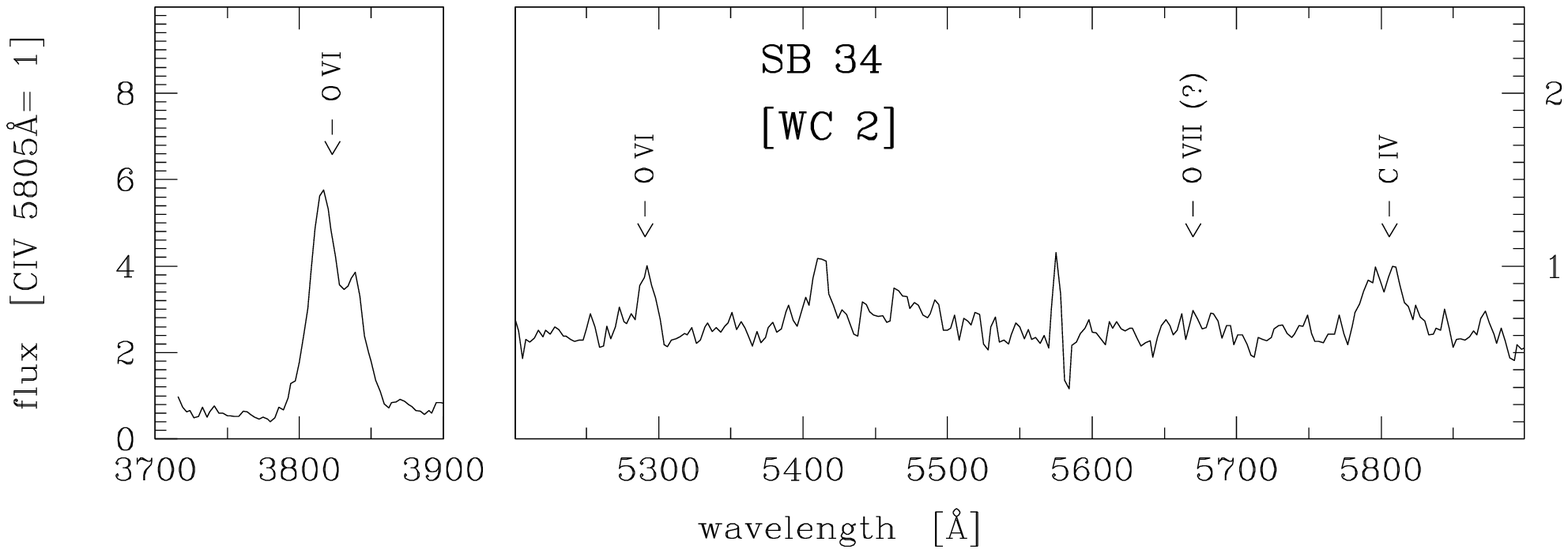}}
\resizebox{0.81\hsize}{!}{\includegraphics{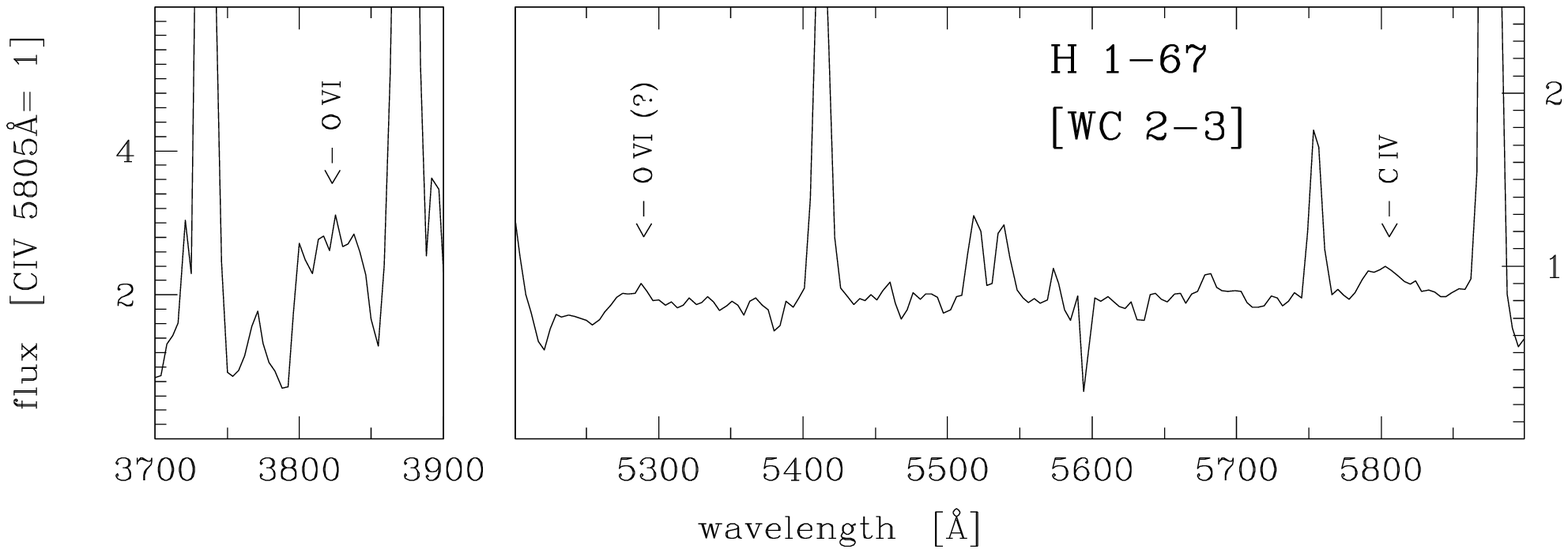}}
\resizebox{0.81\hsize}{!}{\includegraphics{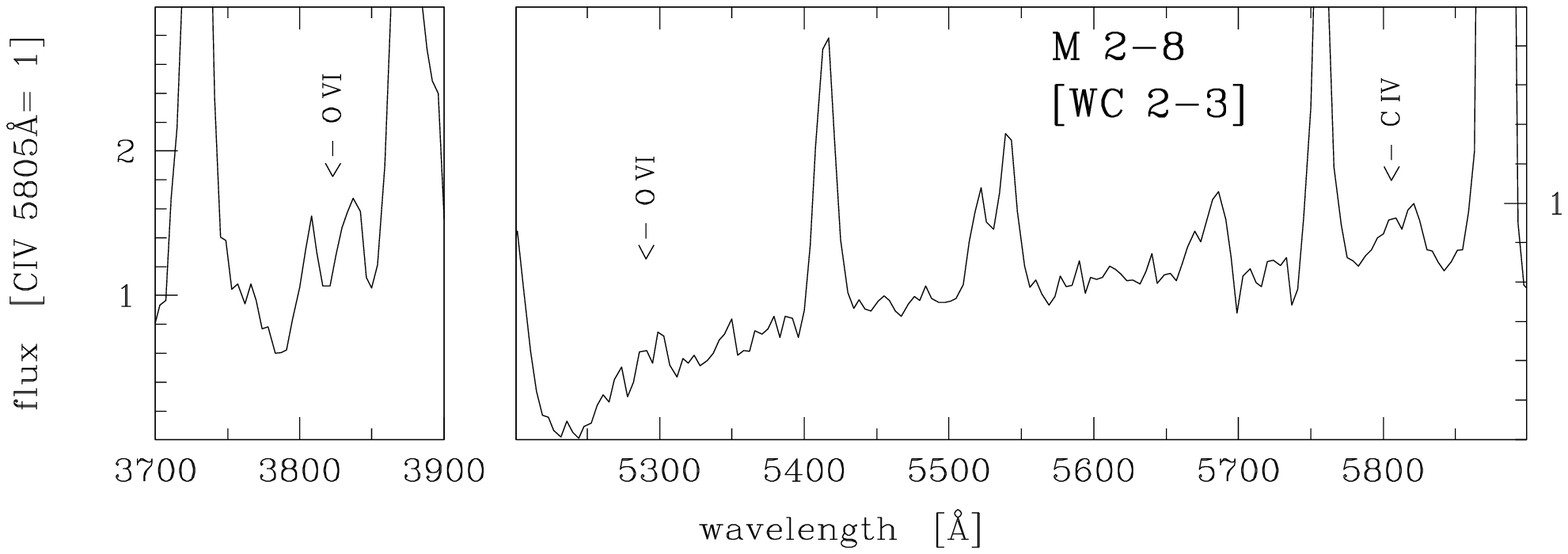}}
\resizebox{0.81\hsize}{!}{\includegraphics{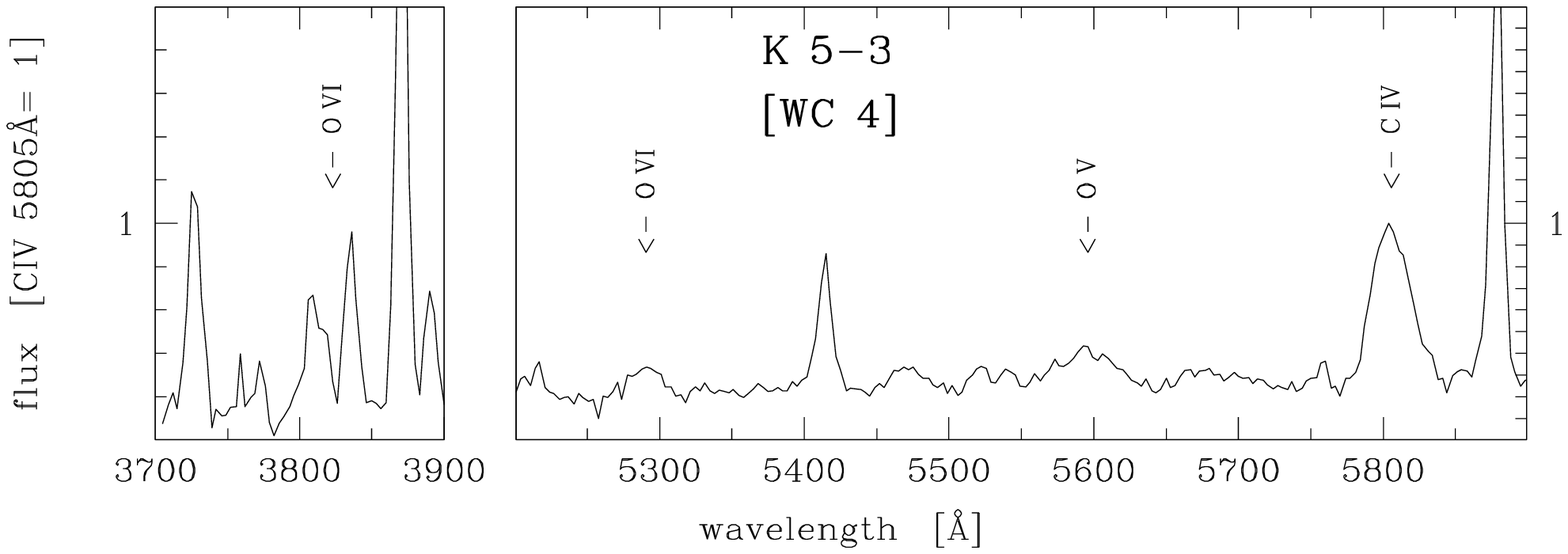}}
\resizebox{0.81\hsize}{!}{\includegraphics{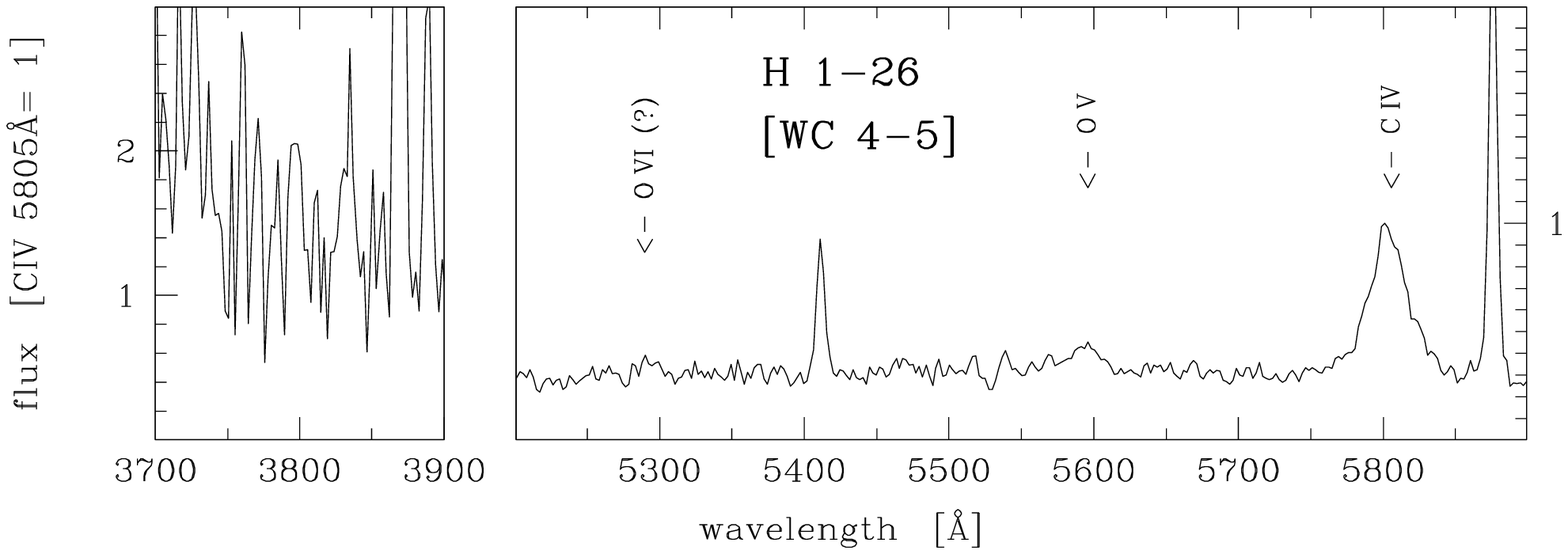}}
\resizebox{0.81\hsize}{!}{\includegraphics{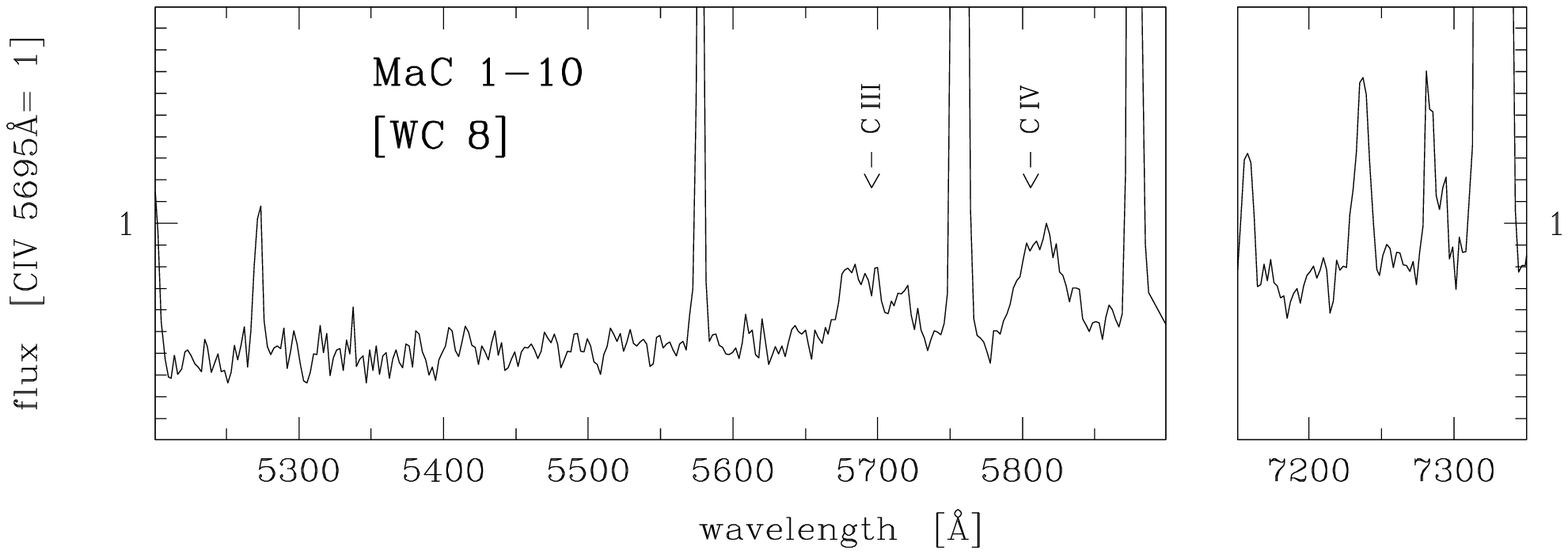}}
\resizebox{0.81\hsize}{!}{\includegraphics{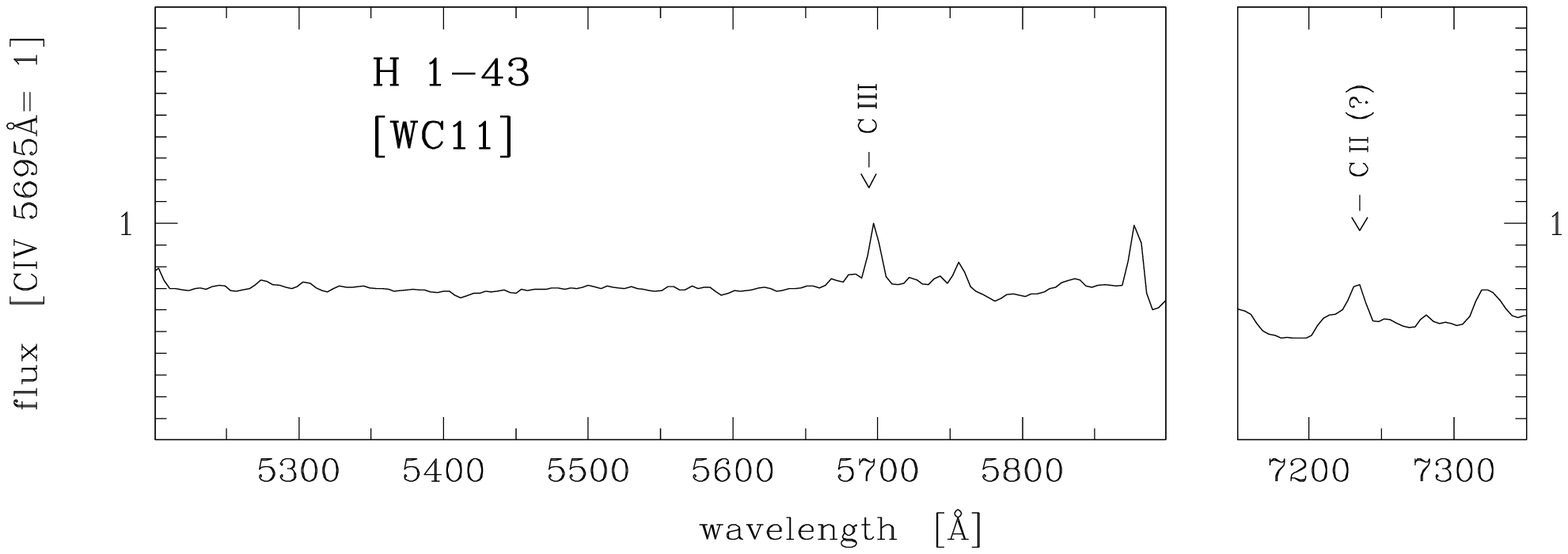}}
\resizebox{0.81\hsize}{!}{\includegraphics{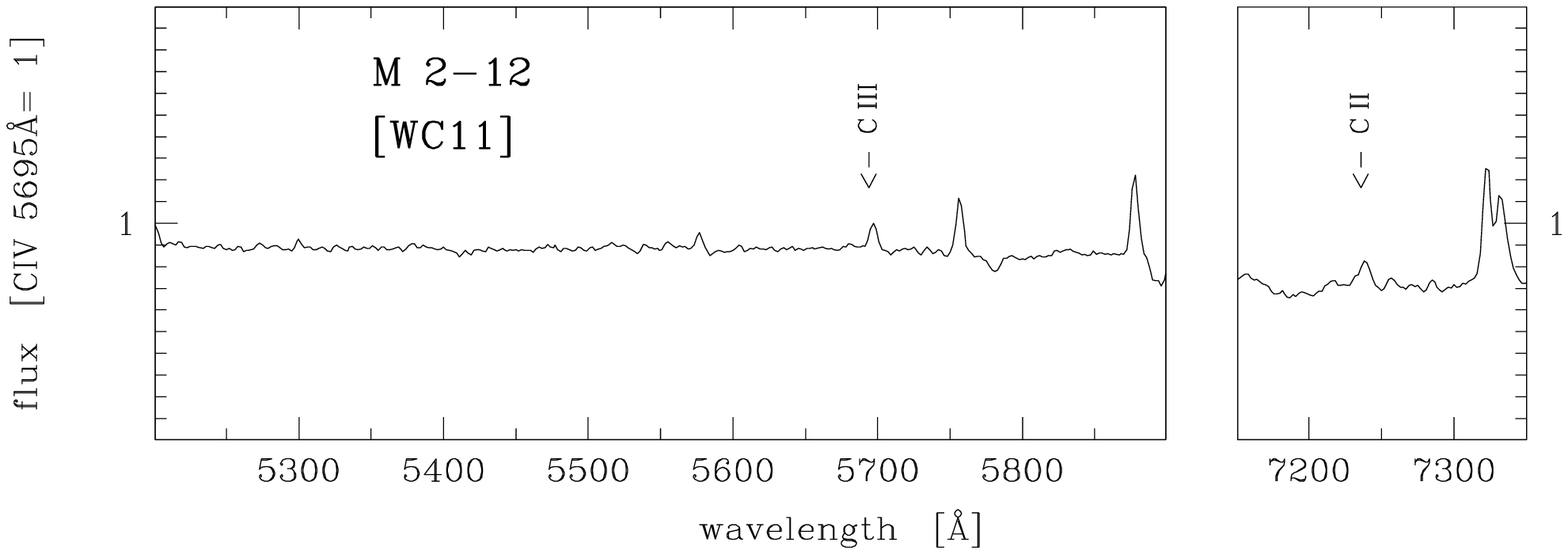}}
\resizebox{0.81\hsize}{!}{\includegraphics{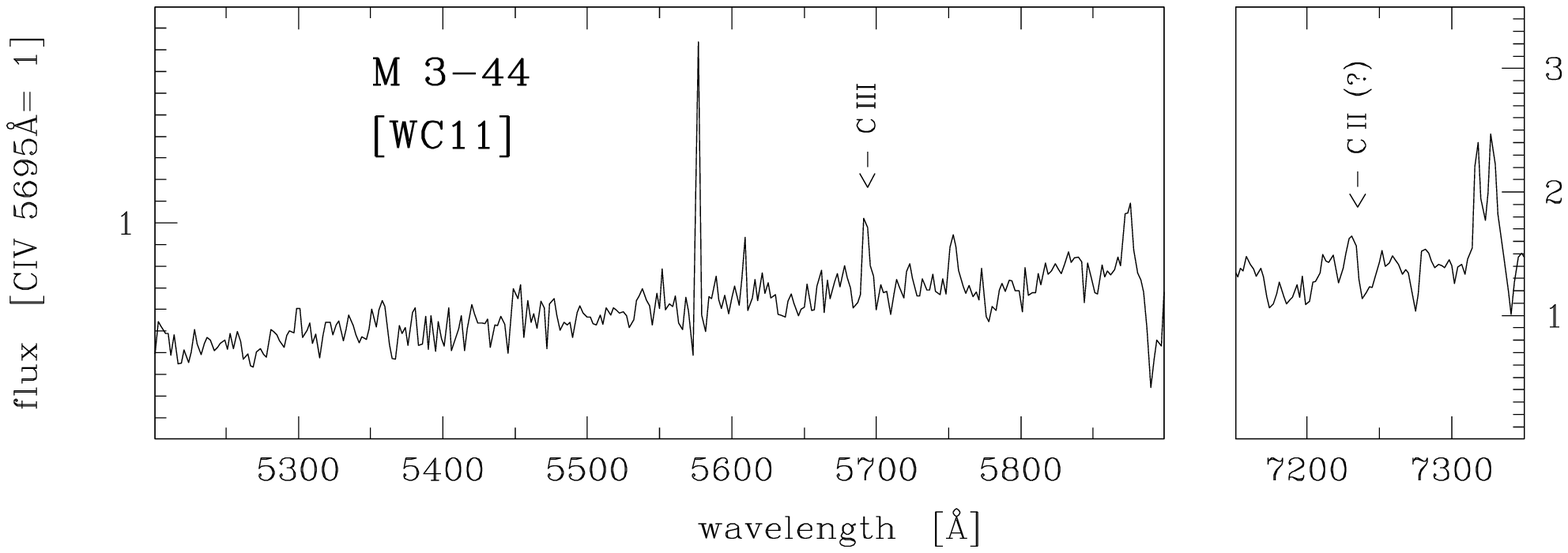}}
\caption[]{
Spectra of the newly discovered [WR]\,PN
}
\label{f5}
\end{figure}

Fig.~\ref{f5} presents the spectra of all the objects we identifed as
[WR]\,PNe in samples G, B, K and E.

\subsection{[WR]\,PNe spectral classification}

The classification criteria we used followed the classical scheme (van der
Hucht et al. \cite{vdHCL81}, M\'endez \& Niemela \cite{MN82}, Hu \& Bibo
\cite{HB90}) which is based solely on relative intensities of selected
lines: C~{\sc iv} $\lambda$ 5805, C~{\sc iii} $\lambda$5695, C~{\sc ii}
$\lambda$5663, O~{\sc vii} $\lambda$5670, O~{\sc vi} $\lambda$5290 and
O~{\sc v} $\lambda$5595. In addition, the presence of O~{\sc vi}
$\lambda$3822 and C~{\sc ii} $\lambda$7235 was also checked. The O~{\sc vi}
$\lambda$3822 line is expected to be several times brighter than the basic
C~{\sc iv} $\lambda$5805 line in early [WC] spectral types, therefore even
if subject to large extinction it may be more easily detectable than the
C~{\sc iv} $\lambda$5805 line. On the other hand C~{\sc ii} $\lambda$7235 is
expected to be the brightest line in the late [WC] types (Acker \& Neiner
\cite{AN03}) and replaces the hardly detectable C~{\sc ii} $\lambda$5663 line
originally proposed by Hu \& Bibo (\cite{HB90}).

Recently two refined classifications schemes have been proposed for
Wolf-Rayet central stars of planetary nebulae based on quantitative criteria
(Crowther et al. \cite{CdMB98}, Acker \& Neiner \cite{AN03}). The first one
imposes identical criteria of classification for massive population I WC
stars and central stars of planetary nebulae but requires good quality
spectra in order to derive reliable equivalent width measurements of stellar
emission lines, especially if the nebular continuum is dominating. The
second one is, in principle, better adapted to the CSPN specificity.
Unfortunately, the primary criteria are not clearly defined. In addition,
the ranges in line ratios defining the classes show strong discontinuities.
Finally, the classification of early-types is not independent of the
chemical composition. For all these reasons we decided to continue to use in
this work the classical criteria. These criteria are more suited to the
quality of our observations and can be directly compared to most of previous
classification works.

The results of our spectral classification of [WR] stars are presented in
columns (7) and (8) of Table 4. Column (7) gives the [WR] type. Column (8)
describes the details of our spectra indicating this type or gives the
references of the papers were the classification was published. For
reference purposes, column (9) gives the spectral class assigned by Acker \&
Neiner (\cite{AN03}) to the [WR]\,PNe known at that time. We have tried to
use the classification scheme of Acker \& Neiner (\cite{AN03}) also to our
newly discovered [WR]\,PNe. The results are given in the same column in
parenthesis.

As can be seen from Table 4, a large proportion of the [WR] stars we
discovered can be best described as late [WC11] type using the long-used,
classical classification scheme adopted here. Their spectra look as expected
for a very late WC type at a large distance. They are characterized by the
presence of the narrow (FWHM $<$ 5\,{\AA}) C~{\sc iii} $\lambda$5695 line
and the absence of the C~{\sc iv} $\lambda$5805 line. Therefore they fulfil
one of the criteria to classify them as [WC11] (Hu \& Bibo \cite{HB90}). For
some objects, especially the ones belonging to sample C, a classically
definied [WC10] type cannot be excluded, since better spectra might reveal
the presence of weak C~{\sc iv} lines. Interestingly, the C~{\sc ii}
$\lambda$7235 line was definitely seen in only one of the spectra we
examined, while it was proposed that the [WC11] type should be dominated by
both C~{\sc iii} and C~{\sc ii} lines (Hu \& Bibo \cite{HB90}). Again, this
might be because the C~{\sc ii} lines are too weak to be detected in our
spectra therefore we regard this classification criteria as less important
and imprecise.

Note that even the new classification schemes would require some
clarification as far as the [WC11] class is concerned. In both papers
(Crowther et al. \cite{CdMB98} and Acker \& Neiner \cite{AN03}), only one
object, \object{K 2-16}, is classified as [WC11].  The defining criteria of
the [WC11] class are obtained by extrapolation of the properties from
adjacent classes and it is therefore not certain what intensity of C~{\sc
ii} lines should be expected. It must also be mentioned that \object{K 2-16}
has a rather unusual spectrum showing C~{\sc iv} $\lambda$5805 and O~{\sc v}
$\lambda$5595 in absorption and surprisingly also some photospheric lines
not expected to be present in [WR] stars (Crowther et al. \cite{CdMB98},
de~Ara\'ujo et al. \cite{dAMP02}). It was proposed by Pe\~na et al.
(\cite{PSM01}) that this object may have undergone a late helium-shell
flash, so that it did not follow the common evolutionary path in which the
[WR] phenomenon appears just after the AGB phase (see G\'orny \& Tylenda
\cite{GT00}).  This is supported by the distinct properties observed for
\object{K 2-16} in the infrared domain (G\'orny et al. \cite{GSS01},
Szczerba et al. \cite{SGS01}, Szczerba et al. \cite{SVG03}).

\subsection{Statistics of [WR] types}

Having determined which of the [WR]\,PNe  are most likely physically linked
to the bulge (see Sect. 2.2), we can compare the distributions of [WC] types
in the bulge and in the disk.

\begin{figure}
\resizebox{0.9\hsize}{!}{\includegraphics{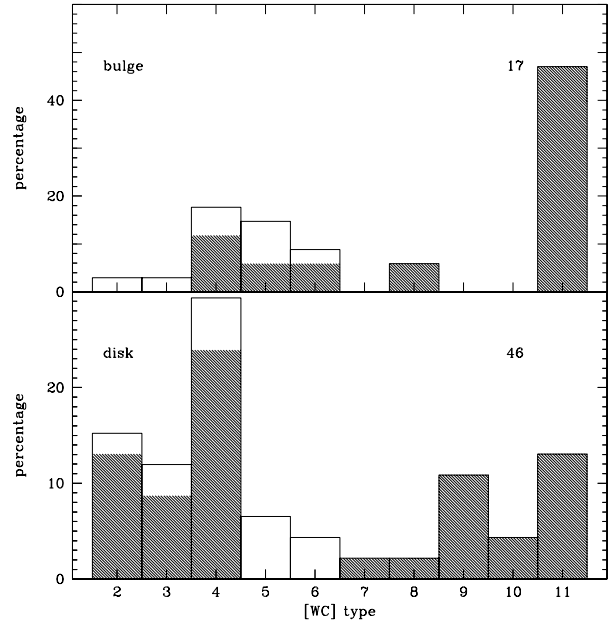}}
\caption[]{
The histograms of [WC] types for  [WR]\,PNe in the Galactic bulge as
defined in Sect. 2.2 (top) and in the Galactic disk (bottom) for all
known [WR]\,PNe in the Galaxy including our newly discovered sources. The
total number of objects used to construct each histogram is given in
the upper right corner. The empty parts of the bars correspond to [WC 2-3],
[WC 3-4] [WC 4-5], [WC 4-6] and [WC 5-6] types, distributed over two
neighbouring bins.
}
\label{f6}
\end{figure}

Figure~\ref{f6} shows the histograms of [WC] types for [WR]\,PNe pertaining
to the bulge (top) and for [WR]\,PNe pertaining to the disk population
(bottom). If we consider only objects from our "merged sample" (described in
Sect. 2.2.) it is remarquable that all the three [WR]\,PNe in sample "d" are
of type earlier than [WC6] while there are only 5 such objects among 14 in
sample "b". Since the statistics for our sample "d" are extremely low, we
consider in the Fig.~\ref{f6} all the known [WR] PNe by merging the ones
from Table 4 with those collected in the litterature by G\'orny \& Tylenda
(\cite{GT00}). The top histogram concerns [WR]\,PNe of the Galactic bulge
population, the bottom one shows all the known [WR]\,PNe pertaining to the
entire Galactic disk. The litterature data add to our "merged sample" only 3
[WR]\,PNe in the bulge, but increases the disk sample from 3 to 46 (as most
of the known [WR]\,PNe lie outside the direction of the galactic bulge). 
The difference between the bulge and disk histograms is highly significant.

First, we find  a much larger proportion of [WR] stars later than [WC10] in
the entire bulge sample (47\%) than in the disk sample (17\%). Second, the
[WR]\,PNe in the direction of the Galactic bulge are underrepresented in
spectral types [WC2] and [WC3] with respect to Galactic disk [WR]\,PNe. We
have found only one such case (M 2-8) (i.e. 6\%) in the Galactic bulge while
there are 26\% of objects in the disk sample. Note, finally, that the
proportion of intermediate classes ([WC5] to [WC6]) is significantly larger
in the bulge sample (22\%) than in the disk sample (11\%). We thus confirm
the claims by G\'orny (\cite{Go01}) which were based on the analysis of much
smaller samples.

\begin{figure}
\resizebox{0.9\hsize}{!}{\includegraphics{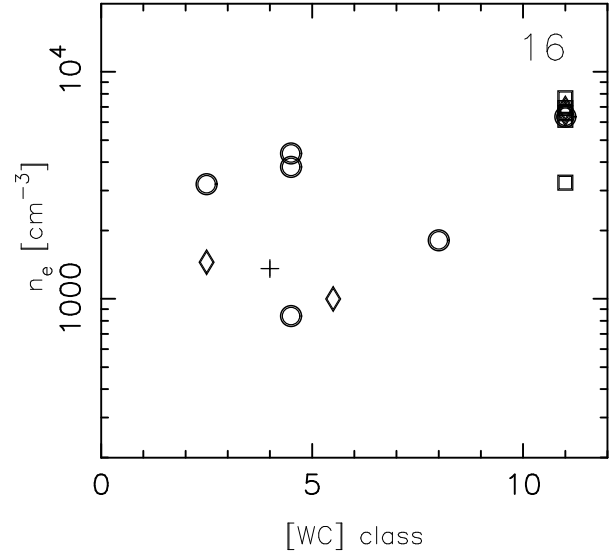}}
\caption[]{
The nebular density deduced from [S~{\sc ii}] $\lambda$6731/6717 as a
function of the [WC] type for all the newly discovered [WR]\,PNe. Objects
pertaining to different samples are marked with the same symbols as in
Fig.~\ref{f1}.
}
\label{f7}
\end{figure}

It has been suggested  (e.g. G\'orny \& Tylenda \cite{GT00}, Pe\~na et al. 
\cite{PSM01}) that the [WC] sequence is an evolutionary sequence, with most
of the late-type [WC] being surrounded by high density PNe and most of the
early-type [WC] being surrounded by low density PNe. In Fig.~\ref{f7}, we
show the nebular density deduced from [S~{\sc ii}] $\lambda$6731/6717 as a
function of the [WC] type for all the newly discovered [WR]\,PNe. The same
trend is found as previously, well in line with the hypothesis that most
[WR]\,PNe draw their evolution directly from the AGB.

% figure 6

% figure 7

% figure 8
\begin{figure}
\resizebox{0.9\hsize}{!}{\includegraphics{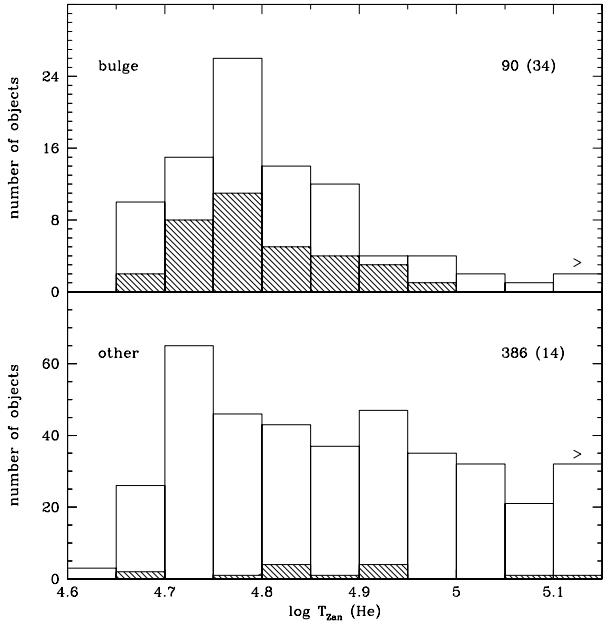}}
\caption[]{
The histogram of Zanstra He~{\sc ii} temperature for the PNe catalogued in
Acker et al. (\cite{AOS92}) that pertain to the Galactic bulge (upper panel)
and for the remaining PNe (lower panel). The number of objects used to
construct the plot is given in the top right of each panel. The dashed bars
stand for objects in common with our G, C or E samples and their number is
given in parenthesis.
}
\label{f8}
\end{figure}

\subsection{Selection effects}

It may be surprising that most of the [WR]\,PNe we discovered are of late
types, since one would expect the earliest [WC] types to be more easily
detectable than the latest types, because the stellar features are both
wider and stronger. However, before interpreting our findings in terms of
stellar evolution or stellar masses, one should examine the selection
effects more closely.

First of all, in the bulge, where the conditions of detection of PNe are
difficult due to high extinction and crowding, there is a discrimination
against detection of PNe with very hot central stars, because such nebulae
are less luminous and more diffuse. This probably explains, at least in
part, why the proportion of PNe with high effective temperatures is higher
in the disk than in the bulge, as seen in Fig.~\ref{f8} which compares the
observed distribution of of He~{\sc ii} Zanstra temperatures,
$T_{\rm{Zan}}$(He~{\sc ii}), in the bulge (upper panel) and in the disk
(lower panel). To construct this figure, we have used the criteria described
in Sect.~2.2 to divide the Strasbourg-ESO catalogue of Galactic PNe (Acker
et al. \cite{AOS92}) into one subsample of 263 objects that most likely
belong to the bulge, and one subsample of 878 objects which, for their vast
majority, belong to the disk. We computed $T_{\rm{Zan}}$(He~{\sc ii}) for
all the PNe with central stars hot enough to ionise He$^{+}$ and for which
the necessary data could be found in Acker et al. (\cite{AOS92}) and Tylenda
et al. (\cite{TSA94}), obtaining values of He~{\sc ii} Zanstra temperatures
for 90 PNe in the bulge and 386 in the disk. The error bar on
$T_{\rm{Zan}}$(He~{\sc ii}) is typically of 0.05 in the log. The dashed
bars in Fig.~\ref{f8} stand for the objects present in our G, C or E
subsamples. We see that, in the bulge, the number of PNe with sufficiently
hot central stars (log~$T_{\rm{Zan}}$(He~{\sc ii}) $>$4.9) is very small,
making the discovery of an early type [WC] star very unlikely. It is also
interesting to note that the proportion of objects in the range
4.7$>$log~$T_{\rm{Zan}}$(He~{\sc ii})$>$4.85 (corresponding to the
temperatures of intermediate [WC] types) is in fact similar in both groups
of PNe: 21\% (55/263) for the Galactic bulge objects and 18\% (154/878) for
the remaining PNe.

% figure 9
\begin{figure}
\resizebox{0.9\hsize}{!}{\includegraphics{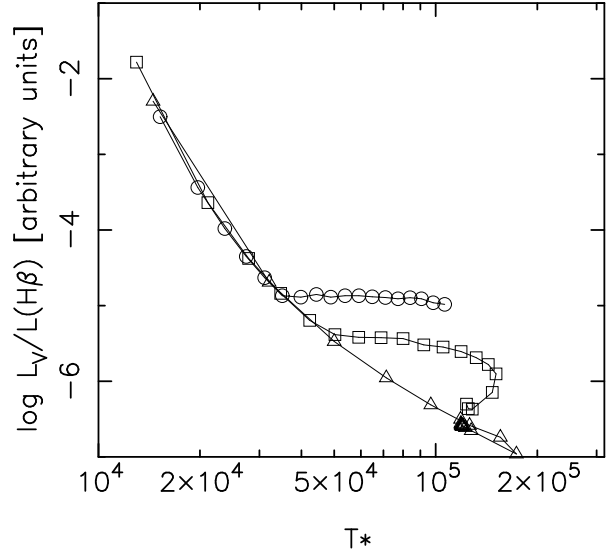}}
\caption[]{
The ratio $L_{V}$/$L$(H$\beta$) as a function of the star effective
temperature, $T*$, for a series of simple models of expanding planetary
nebulae around central stars of different masses (see text). Symbols are
placed every 500~yr. Circles: $M_{\star}$=0.58~M$_{\odot}$; squares:
$M_{\star}$ =0.60~M$_{\odot}$; triangles: $M_{\star}$ =0.62~M$_{\odot}$.
}
\label{f9}
\end{figure}

% figure 10
\begin{figure}
\resizebox{0.9\hsize}{!}{\includegraphics{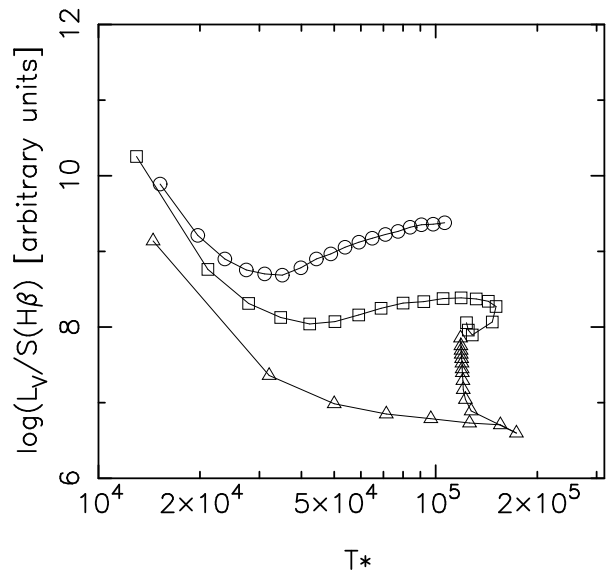}}
\caption[]{
The ratio $L_{V}$/$S$(H$\beta$) as a function of $T*$. Same models and
same presenation as in Fig.~\ref{f9}.
}
\label{f10}
\end{figure}

Next, one must consider the condition of detection of the stellar emission
lines.

PNe in the bulge have small angular dimensions, so that a large fraction of
their total nebular flux is recorded in their spectra. Therefore the
detection of [WR]\,PNe is possible only for objects in which the luminosity
of the star in the V band ($L_V$) is sufficiently large with respect to the
total nebular continuum emission (which is proportional to the luminosity in
H$\beta$). For illustrative purposes, we show in Fig.~\ref{f9} the ratio
$L_{V}$/$L$(H$\beta$) as a function of the star effective temperature,
$T*$, for a series of simple models of expanding planetary nebulae around
central stars of different masses (the central stars evolve according to an
interpolation of Bl\"ocker \cite{Bl95} evolutionary tracks, the nebulae are
homogeneous with a total mass of 0.3~M$_{\odot}$, and the expansion velocity of the
outer rim is taken to be 20~km~s$^{-1}$). This corresponds to the extreme case where
the entire nebula is covered by the region of the slit where the spectrum is
extracted). This figure shows that the most favourable cases for detection
of [WR] features in nebulae of small angular dimentions are those
corresponding to the lowest stellar temperatures, i.e. to the latest
spectral types, despite the fact that stellar emission lines are stronger in
early-types.

On the contrary, many of the PNe known in the Galactic disk are relatively
close-by, and have large angular dimensions, so only a small fraction of the
nebular emission is recorded in the spectrograph slit. Then, a [WR] star
will be preferentially detected in PNe with large $L_{V}$/$S$(H$\beta$)
(where $S$(H$\beta$) is the nebular surface brightness in H$\beta$).
Fig.~\ref{f10} shows this ratio as a function of $T*$ for the same models as
Fig.~\ref{f9}. We see that, the conditions of detectability of [WR] features
are more complicated. In the early phases of evolution, the
$L_{V}$/$S$(H$\beta$) ratio is large, but this regime does not last long. At
intermediate values of $T*$, the conditions of detectability of stellar
features are less favourable. Finally, at high values of $T*$, the
$L_{V}$/$S$(H$\beta$) ratio increases due to the expansion of the nebula,
and the star becomes easier to observe, especially if the object is
close-by.

Note that, especially for close-by nebulae with high temperature central
stars, stellar features are much easier to detect for low than for high
central star masses.

The distribution of [WR] stars among spectral classes is thus highly
dependent on the selection effects just described. In addition, one must
recall that the quality of the spectra used for the discovery and
classification of [WR]\,PNe is very heterogeneous. For example, the quality
of our new observations presented in Sect. 2.1, specially aimed at
discovering new [WR]\,PNe is much higher than that of the survey on which
was based the seminal paper by Tylenda et al. (\cite{TAS93}).  All this
badly hampers any attempt to use [WR] types statistics to study the
populations of the central stars and their evolution.

\subsection{Do [WR]\,PNe occur more frequently in the Galactic bulge
than in the disk?}

We now compare the global characteristics of the  [WR]\,PNe in our "merged
sample" with those of the remaining PNe in this sample.  Fig.~\ref{f1} shows
that the [WR]\,PNe are more concentrated than the non [WR]\,PNe in the
($l_{\rm{II}}$, $b_{\rm{II}}$) diagram and have also larger H$\beta$ fluxes.
It is interesting that the [WR]\,PNe that lie close to the center of the
Galaxy are of late [WC11] type.

Among the objects whose spectra we examined, there are 15 [WR]\,PNe among a
total of 94 objects in the $b$ subsample and 3 [WR]\,PNe among a total of 52
objects in the $d$ subsample. Thus [WR]\,PNe appear preferentially in the
bulge population than in the disk population, at least in the direction of
the Galactic bulge. Does this mean that the {\emph{real}} proportion of
[WR]\,PNe is larger in the former than in the latter? Since {\emph{all}} the
spectra from our "merged sample" have been obtained in the same conditions
and have been examined in a similar way for the possible presence of [WR]
stars, the only effect to consider is whether it is easier to recognize a
[WR] star in bulge or in disk PNe. A priori, it is easier to recognize a
[WR] in a nebula of large angular dimensions and relatively close-by.
Therefore, we conclude that the true proportion of [WR]\,PNe is
significantly larger in the bulge than in the disk. Keeping in mind the
considerations in Sect. 4.4, this could indicate that, in the direction of
the Galactic bulge, the central star masses are likely smaller, on average,
in the bulge than in the disk.

\section{Abundances}

\subsection{Oxygen abundances: bulge versus disk}

Cuisinier et al. (\cite{CMK00}) compared the oxygen abundances of bulge and
disk PNe (their Fig. 5) by considering their own sample for the bulge (which
indeed strictly obeys our criteria for sample $b$) with data taken from
Maciel \& K\"oppen (\cite{MK94}) for the Galactic disk. They found that the
oxygen abundances in bulge PNe are comparable to those in disk PNe, with a
slightly higher proportion of high abundances in the bulge. However, the
distributions of the oxygen abundances in PNe in the entire Galactic disk
shows important gradients (Maciel \& K\"oppen \cite{MK94}, Maciel \& Quireza
\cite{MQ99}) so that such a comparison is of limited utility. With our
merged data set, we can do somewhat better, because our sample $d$ contains
only objects in the direction of the Galactic bulge, and is likely dominated
by objects located between the Sun and the Galactic center, with a larger
proportion of objects at small galactocentric distances.

% figure 11
\begin{figure}
\resizebox{0.9\hsize}{!}{\includegraphics{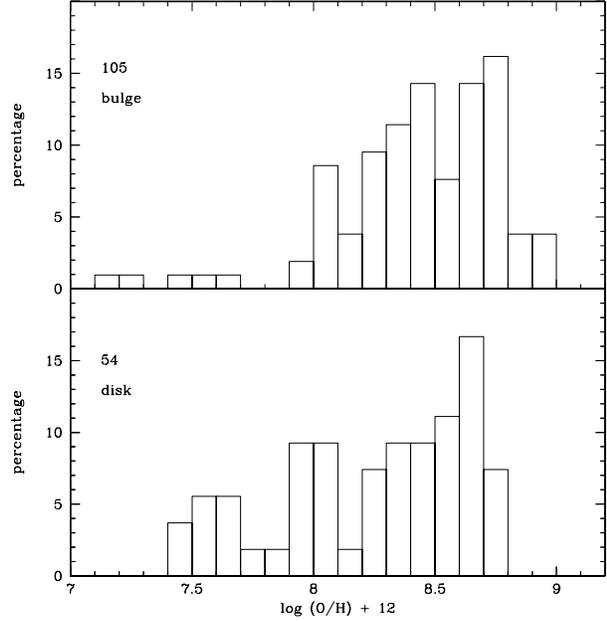}}
\caption[]{
The histogram of the oxygen abundances we derived for the objects of our
merged sample that pertain to the bulge subsample $b$ (top panel) and
for the objects that pertain to the disk subsample $d$ (bottom panel)
as defined in Sect. 2.2.
}
\label{f11}
\end{figure}

In Fig.~\ref{f11}, we show the histograms of the oxygen abundances for our
sample $b$ and $d$. One can see that the distribution is significantly
narrower and more skewed to high O/H values in sample $b$. Non parametric
statistical tests confirm this impression.  The median value, and the 25 and
75 percentiles are 8.45 (8.21, 8.68) for sample $b$ while they are 8.32
(7.93, 8.59) for sample $d$. The above considerations concern the nominal
abundances derived from the observed line intensities. Errors in abundance
derivations were discussed in Sect. 3.2. They are not expected to change our
conclusions, except perhaps as regards the PNe which were found to have very
low abundances.  However, of the 5 PNe with a nominal value of log O/H +12
smaller than 7.5 in the entire sample, 2 have abundance with better than
0.2~dex accuracy as seen in Fig.~\ref{f4}, and for the three remaining ones,
two pertain to sample $b$ and one in sample $d$. In addition, we should
consider that, within our entire sample of observations, there are 4 objects
in sample $b$ and 1 in sample $d$ for which the abundances could not be
derived because the electron temperature is unknown. The relative proportion
of $b$ and $d$ objects among these 4 objects, compared to a total of 105
objects in sample $b$ and 54 objects in sample $d$ implies that our
conclusions are not likely to be affected by errors in the abundances.
Clearly, better data on a larger sample would be needed to confirm this.

From the compilation of Maciel \& Quireza (\cite{MQ99}), which concerns PNe
at Galactocentric distances between 3 and 14\,kpc, the oxygen abundance
steadily increases inwards, reaching log O/H +12 of about 8.7 at the
Galactocentric distance of the Sun and about 9 at 3\,kpc, with a dispersion
of about $\pm$ .2\,dex. Comparing these values with the histogram shown in
Fig.~\ref{f11} for our $d$ sample (lower panel), we conclude that O/H levels
off and probably even decreases in the most internal parts of the Galactic
disk. This is well in line with the findings of Smartt et al. (\cite{SVD01})
based on abundance determination of four B-type stars.

As concerns the bulge, the oxygen abundance distribution we find has roughly
the same shape as that found by Zoccali et al. (\cite{ZRO03}) for the
metallicity distribution of bulge giants (with a sharp cut-off at the high
abundance end and a long tail towards low abundances), but it is narrower by
roughly 0.3~dex.

It is remarkable also that the median oxygen abundance in sample $b$ is
larger by about 0.2\,dex than that of our sample $d$.

% figure 12
\begin{figure}
\resizebox{0.9\hsize}{!}{\includegraphics{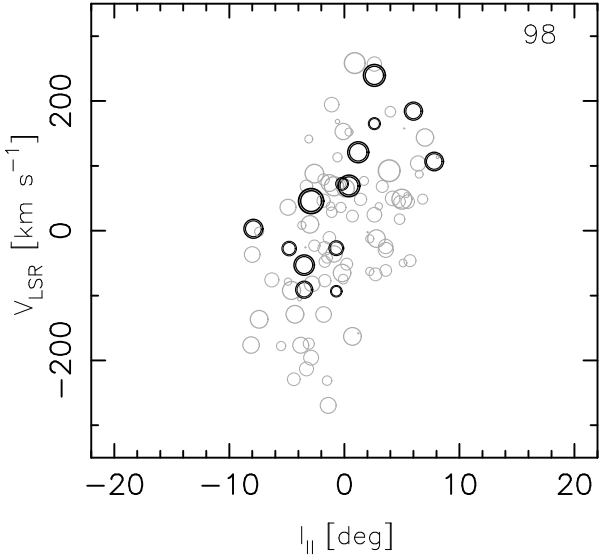}}
\caption[]{
The ($V_{\rm{lsr}}$, $l_{\rm{II}}$) diagram  for samples $b$ only.
The radius of the circle is proportional to O/H.
}
\label{f12}
\end{figure}

% figure 13
\begin{figure}
\resizebox{0.9\hsize}{!}{\includegraphics{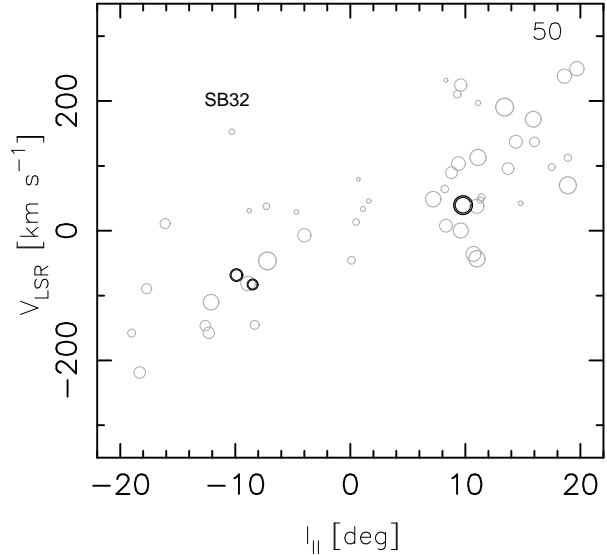}}
\caption[]{
The ($V_{\rm{lsr}}$, $l_{\rm{II}}$) diagram  for samples $d$ only. The
radius of the circle is proportional to O/H. The object \object{SB 32}
(\object{PN G 349.7-09.1}), which possibly belongs to the halo (see Sect.
5.1) has been labelled.
}
\label{f13}
\end{figure}

We also examined whether oxygen abundances show any  trend with some other
properties. The most interesting results are seen in Figs.~\ref{f12}
and~\ref{f13}, where we show again the ($V_{\rm{lsr}}$, $l_{\rm{II}}$)
diagram for samples $b$ and $d$ separately, this time representing the
nebulae by circles whose surface is proportional to O/H. Fig.~\ref{f13}
displays one important feature. While most of the objects are roughly
consistent with rotation of the Galactic disk, one object is clearly an
outlier. This is \object{SB 32} (\object{PN G 349.7-09.1}), which has log
O/H + 12 = 7.77 (+0.47, -0.24). Both its position in the ($V_{\rm{lsr}}$,
$l_{\rm{II}}$) diagram and its  low oxygen abundance indicate that this PN
likely belongs to the Galactic halo and not to the disk.

In Fig.~\ref{f12}, which concerns sample $b$, we may note that the PNe with
smaller oxygen abundances tend to have smaller radial velocities.

\subsection{Oxygen abundances: [WR]\,PNe versus other PNe in the bulge}

% figure 14
\begin{figure}
\resizebox{0.9\hsize}{!}{\includegraphics{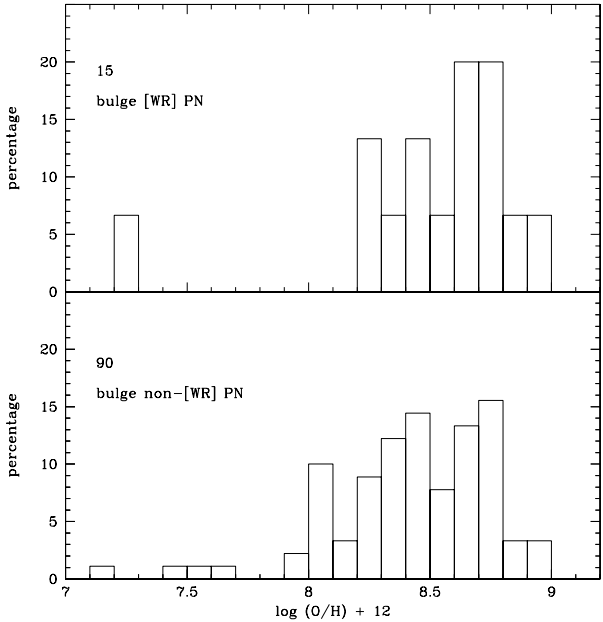}}
\caption[]{
The histogram of the oxygen abundances we derived for the [WR]\,PNe of our
merged sample (top panel) and for the non-[WR]\,PNe (bottom panel) that
pertain to the bulge subsample $b$ as defined in Sect. 2.2.
}
\label{f14}
\end{figure}

When comparing the abundances of [WR]\,PNe and non[WR]\,PNe, G\'orny \&
Stasi\'nska (\cite{GS95}) refrained from comparing oxygen abundances, which
are strongly affected by the location of the PN in the Galaxy. With our
present work, we are in a much more favourable situation. We have a sample
of PNe in the bulge out of which was drawn a sample of [WR]\,PNe.
Fig.~\ref{f14} shows in the top panel the histogram of the oxygen abundances
for the [WR]\,PNe in our sample $b$ while in the bottom panel the same is
shown for the remaining PNe in this sample. Although the sample of [WR]\,PNe
is small (15 objects), the distributions are remarkably similar and
undistinguishable by non parametric statistical tests.

We thus conclude that the oxygen abundances in [WR] stars are not
significantly affected by nucleosynthesis and mixing in the progenitors.
This implies that [WR]\,PNe, similarly to ordinary PNe, can be used to
determine the oxygen abundance of the media out of which the progenitors
were born.

\section{N/O ratios}

% figure 15
\begin{figure}
\resizebox{0.9\hsize}{!}{\includegraphics{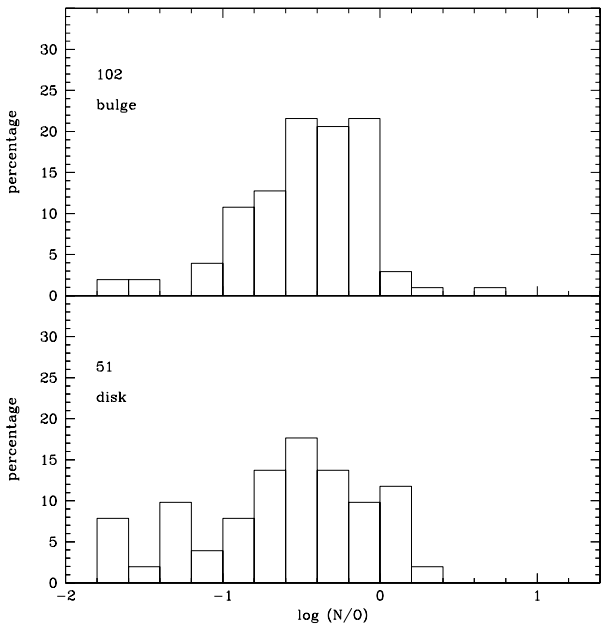}}
\caption[]{
The histogram of the N/O ratios we derived for the PNe of our merged sample
that pertain to the bulge subsample $b$ (top panel) and for the PNe that
pertain to the disk subsample $d$ (bottom panel) as defined in Sect. 2.2.
}
\label{f15}
\end{figure}

% figure 16
\begin{figure}
\resizebox{0.9\hsize}{!}{\includegraphics{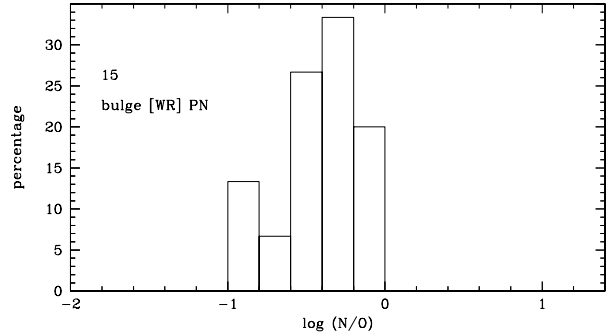}}
\caption[]{
The histogram of N/O ratios derived for the [WR]\,PNe of our merged sample
that pertain to the bulge subsample $b$ as defined in Sect. 2.2.
}
\label{f16}
\end{figure}

Fig.~\ref{f15} show the histograms of the N/O ratios in samples $b$ (top
panel) and $d$ (bottom panel) respectively. The distributions are seen to be
similar and they are undistinguishable by non parametric statistical tests.

Fig.~\ref{f16} shows the histogram of N/O in the [WR]\,PNe of sample $b$.
Although also in this case the number of objects is small (15), again we can
conclude that the distribution is indistinguishable, statistically, from the
previous ones.

We thus conclude that, although N is known to be produced and dredged-up by
the PNe progenitors, and that the nitrogen enrichment is expected to by a
function of the progenitor's mass and of the mass loss rate (Marigo
\cite{Ma01}), a statistical study such as the one we present is not able to
give any real clue.

\section{Conclusions}

We have conducted a statistical analysis of planetary nebulae in the
direction of the Galactic bulge, with the double aim of a better
understanding of the global properties of PNe with Wolf-Rayet type central
stars, and of a more refined characterization of the different populations
of PNe that are seen in this direction.

We have performed new observations of 44 PNe in the direction of the
Galactic bulge, and merged them with data from previous studies suitable for
our purposes, constituting a sample of 164 PNe.

We have distinguished, in this merged sample, the PNe most probably
pertaining physically to the Galactic bulge, and the PNe most likely
belonging to the Galactic disk.

We have determined the chemical composition of all the 164 objects in a
coherent way and discussed the uncertainties in the derived abundances.

We have looked for stellar emission features in all the 164 objects. This
search resulted in the discovery of 14 new [WR] stars and 15 new weak
emission line central stars, significantly enlarging the number of emission
line stars known in the direction of the Galactic bulge. We have also
performed a spectral classification of all the new [WR] stars.

Our findings were then discussed with the aim of getting some clues on
[WR]PN production and evolution, on the various populations of planetary
nebulae seen in the direction of the Galactic disk, and on the distribution
of oxygen abundance in the Galaxy as measured by planetary nebulae.
Particular attention was paid to take into account selection effects.

Our main results are the following:

We confirm that the spectral type distribution of [WR] stars is very
different in the bulge and in the disk of the Galaxy. However, we show that
the selection effects, being not the same between bulge and disk PNe, can
explain, at least in part, this observed difference in spectral type
distribution.

We show that the proportion of [WR]\,PNe with respect to the remaining PNe
is significantly larger in the bulge than in the disk. We show that this is
likely an indirect indication that the central star masses are, on average,
smaller in the bulge than in the disk.

There is no difference, in the Galactic bulge, between the oxygen abundances
in [WR]\,PNe and in the remaining PNe. This implies that the oxygen
abundances in [WR] stars are not significantly affected by nucleosynthesis
and mixing in the progenitors. This is the first time that such an inference
can be made, since this is the first time that a significant sample of
[WR]\,PNe and non-[WR]\,PNe belonging exactly to the same Galactic
population is available.

The O/H gradient of the PNe population in the Galactic disk flattens in the
most internal parts of the Galaxy and possibly changes sign, in line with
results from abundance determinations in B-type stars (Smartt et al.
\cite{SVD01}).

The median oxygen abundance in the bulge is larger by 0.2\,dex than that PNe
seen in the direction of the bulge but pertaining to the disk.

The oxygen abundance distribution  for the bulge PNe has a shape similar to
that found by Zoccali et al. (\cite{ZRO03}) for the metallicity distribution
of bulge giants (with a sharp cut-off at high abundance and a long tail at
low abundance), but it is narrower by about 0.3~dex.

The bulge PNe with smaller oxygen abundances tend to have smaller radial
velocities.

Among the PNe seen in the direction of the bulge and pertaining to our
merged sample, \object{SB 32} (\object{PN G 349.7-09.1}) has a location in
the ($V_{\rm{lsr}}$, $l_{\rm{II}}$) diagram and a low oxygen abundance,
indicating that it probably belongs to the Galactic halo population.

Finally, the distributions of N/O are the same in the disk PNe seen in the
direction of the bulge and in PNe pertaining physically to the bulge. They
are also the same in [WR]\,PNe and non[WR]\,PNe pertaining to the bulge.

Some of these findings are in line with what was known before. Others are
rather unexpected. Taken together, and combined with data from other
sources, they should provide interesting constraints on the stellar
populations in the inner parts of the Galaxy, on the Galaxy chemical
evolution and on the causes of the [WR] phenomenon.

\begin{acknowledgements}

S.K.G. acknowledges the "Jumelage Astronomie France-Pologne" as well as the
Polish grant KBN 2.PO3D.017.25 for financial support.  G.S. acknowledges the
"Jumelage Astronomie France-Pologne" and the "PICs Franco-Bresilien" for
financial support. A.V.E. acknowledges FAPESP for his graduate fellowship.

\end{acknowledgements}

%\listofobjects
\end{document}